\documentclass[
showpacs,
superscriptaddress,
nofootinbib,
longbibliography,
3p
]{elsarticle}

\usepackage{amsmath,amssymb,bm,color}
\usepackage{graphicx}
\usepackage[normalem]{ulem}
\usepackage{tikz}
\usepackage[compat=1.1.0]{tikz-feynhand}

\usepackage[
colorlinks=true, 
pdfstartview=FitV, 
linkcolor=blue, 
citecolor=magenta, 
urlcolor=blue, 
bookmarks=true,
bookmarksnumbered=true,
pdftitle={},
pdfauthor={Keisuke Fujii}
]{hyperref}

\newcommand{\pd}[2]{\frac{\partial #1}{\partial #2}}
\newcommand{\dif}[2]{\frac{d #1}{d #2}}

\newcommand{\bx}{\boldsymbol{x}}

\newcommand{\bzero}{\mathbf{0}}
\newcommand{\bp}{\boldsymbol{p}}
\newcommand{\bq}{\boldsymbol{q}}
\newcommand{\bk}{\boldsymbol{k}}

\newcommand{\average}[1]{\langle#1\rangle}

\renewcommand\Im{{\operatorname{Im}}}

\newcommand{\rmbox}{\textrm{box}}

\newcommand{\fp}{\textrm{f-p}}
\newcommand{\pp}{\textrm{p-p}}
\newcommand{\pole}{\textrm{pole}}
\newcommand{\lines}{\textrm{lines}}
\newcommand{\pair}{\textrm{pair}}
\newcommand{\fermion}{\textrm{fermion}}

\newcommand{\calD}{\mathcal{D}}
\newcommand{\calE}{\mathcal{E}}

\newcommand{\calG}{\mathcal{G}}

\newcommand{\calN}{\mathcal{N}}
\newcommand{\calO}{\mathcal{O}}
\newcommand{\calP}{\mathcal{P}}

\newcommand{\calT}{\mathcal{T}}

\newcommand{\dis}{\displaystyle}

\newcommand{\xlabel}[1]{\label{#1}}
\newcommand{\xref}[1]{\ref{#1}}

\begin{document}

\title{Bulk viscosity of resonantly interacting fermions in the quantum virial expansion}

\author[1]{Keisuke Fujii}
\author[1]{Tilman Enss}

\affiliation[1]{
organization={Institute for Theoretical Physics, Heidelberg University},
postcode={D-69120},
city={Heidelberg},
country={Germany}}

\date{\today}

\begin{abstract}
We consider two-component fermions with a zero-range interaction both in two and three dimensions and calculate the bulk viscosity for an arbitrary scattering length in the high-temperature regime.
We evaluate the Kubo formula for the bulk viscosity using an expansion with respect to the fugacity, which acts as a small parameter at high temperatures.
In the zero-frequency limit of the Kubo formula, pinch singularities emerge that reduce the order of the fugacity by one.
These singularities can turn higher-order vertex corrections at nonzero frequencies into the leading order at zero frequency, so that all such contributions have to be resummed.
We present an exact microscopic computation for the bulk viscosity in the high-temperature regime by taking into account these pinch singularities.
For negative scattering lengths, we derive the complete bulk viscosity at second order in fugacity and show that a self-consistent equation to resum the vertex corrections is identical to a linearized kinetic equation.
For positive scattering lengths, a new type of pinch singularity arises for bound pairs.
We show that the pinch singularity for bound pairs leads to a first-order contribution to the bulk viscosity, which is one order lower than that for negative scattering lengths, and that the vertex corrections also provide first-order contributions. We propose a new kinetic equation for bound pairs that derives from a self-consistent equation to resum the vertex corrections.

\end{abstract}

\maketitle
\tableofcontents

\section{Introduction}

The bulk viscosity is one of the fundamental transport coefficients in
hydrodynamics, which determines the dissipation during isotropic
expansion~\cite{Landau-Lifshitz:fluid,Forster:hydro}.
One intriguing property of the bulk viscosity is that it vanishes in conformal
fluids because they can expand isotropically without
dissipation~\cite{Son:2007}.
On the contrary, the nonvanishing bulk viscosity serves as an indicator of
breaking conformal invariance in various
systems~\cite{Schafer:2009,Adams:2012}.
For example, a dilute two-dimensional Fermi gas is scale invariant classically,
but this scale invariance is broken by quantum fluctuations leading to a
quantum scale anomaly, which is theoretically
predicted~\cite{Pitaevskii:1997,Olshanii:2010,Hofmann:2012} and experimentally
observed via 
the isotropic expansion and contraction mode (breathing
mode)~\cite{Holten:2018,Peppler:2018,Murthy:2019}.

In this study, we consider a two-component Fermi gas realized in ultracold
atoms, which provides an ideal ground to study strongly correlated quantum
many-body systems owing to the tunability of the interparticle interaction via
the Feshbach resonance~\cite{Bloch:2008,Giorgini:2008}.
This system has the universality that the interaction effect appears only
through the scattering length, and its universal properties have been actively
explored not only in thermodynamics but also in transport
phenomena~\cite{Zwerger:2012}.
In particular, the system shows nonrelativistic conformality in the unitary
limit~\cite{Mehen:2000,Son:2006,Nishida:2007}.
The vanishing bulk viscosity at unitarity was confirmed
experimentally~\cite{Elliott:2014}.

One simple approach to compute the transport coefficients is the kinetic theory
founded on the quasiparticle
approximation~\cite{Massignan:2005,Bruun:2005,Braby:2010,Bruun:2012,Schafer:2012,Enss:2012,Dusling:2013,Enss:2013,Chafin:2013}.
However, while kinetic theory is capable of predicting the high-temperature
behavior of the shear viscosity and the thermal conductivity of short-range
interacting Fermi gases, the quasiparticle approximation does not capture all
relevant contributions to the bulk viscosity at high
temperatures~\cite{Fujii:2020}.
As another approach in the high-temperature regime, the transport coefficients
can be calculated from the Kubo formula in the quantum virial expansion, whose
expansion parameter is the
fugacity~\cite{Enss:2011,Nishida:2019,Enss:2019,Hofmann:2020,Frank:2020,Maki:2020,Maki:2022,Tanaka:2022}.
In the zero-frequency limit of the Kubo formula, there is the pinch singularity
due to a particle-hole excitation with vanishing energy and momentum
exchange~\cite{Eliashberg:1962,Jeon:1995,Jeon:1996,Hidaka:2011}.
The pinch singularity
reduces the order of the fugacity by one and changes the power counting with respect to the fugacity, so that a resummation is required.
The shear viscosity and the thermal conductivity were computed with an
approximate resummation~\cite{Enss:2011,Nishida:2019,Hofmann:2020,Frank:2020},
and their exact resummation was shown to be equivalent to solving the
linearized Boltzmann equation~\cite{Fujii:2021}.
On the other hand, the bulk viscosity has not yet been calculated taking this singularity into account.
The purpose of this paper is to give an exact microscopic computation of the bulk viscosity in the high-temperature regime by taking into account the pinch singularity.

\begin{figure}[tbp]
 \begin{center}
  \includegraphics[width=100mm]{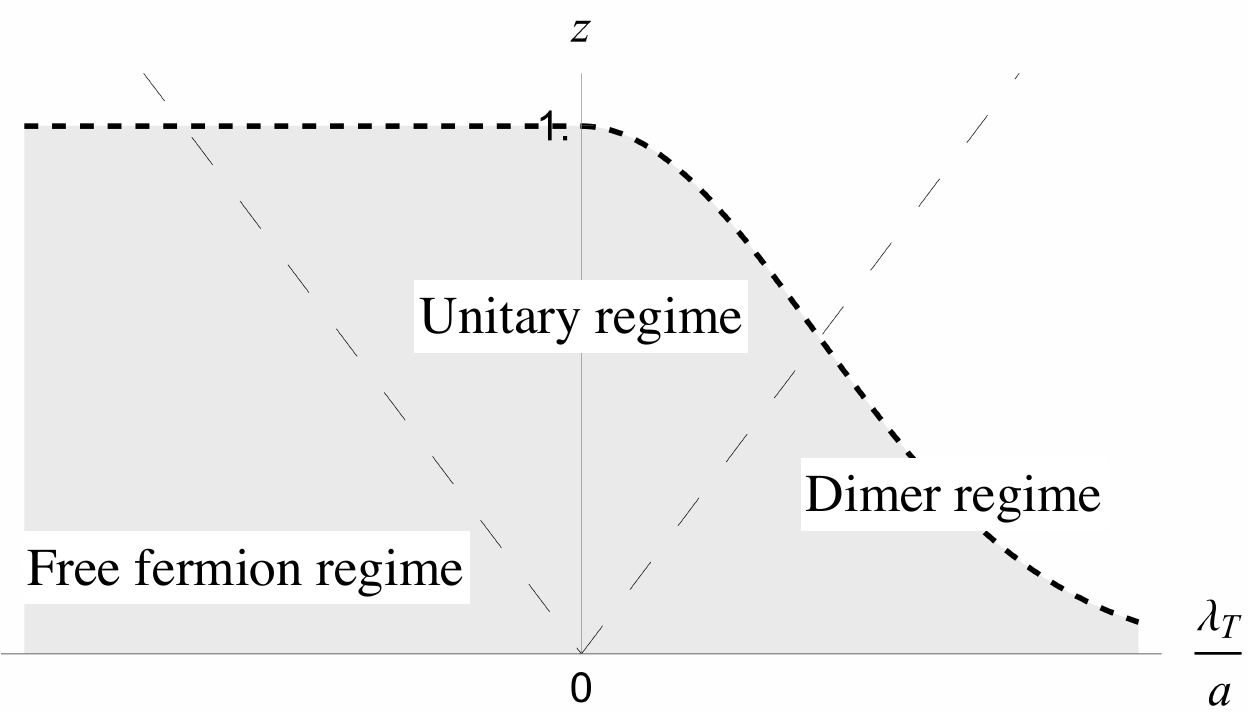}
 \end{center}
\caption{The regime where the quantum virial expansion is valid (the shaded region).
 The horizontal and vertical axis represent the inverse dimensionless scattering length $\lambda_T/a$ and the fugacity $z=e^{\beta\mu}$, respectively.
 At positive scattering length, the expansion is valid when the pair fugacity $z^2e^{\beta/(ma^2)}$ is smaller than the fermion fugacity $z$, which itself should be less than $O(1)$.
 Hence, the applicability of the fermionic virial expansion is limited to $z<e^{-\beta/(ma^2)}$ for $a>0$.
 The free-fermion and dimer limits are often referred to as the BCS (Bardeen--Cooper--Schrieffer) and BEC (Bose--Einstein condensation) limits, but we avoid those terms because we work above the superfluid critical temperature.\label{fig:expansion-regime}}
\end{figure}

\begin{figure}[tp]
\begin{tabular}{cc}
 \begin{minipage}{0.47\hsize}
  \begin{center}
   \includegraphics[width=74mm]{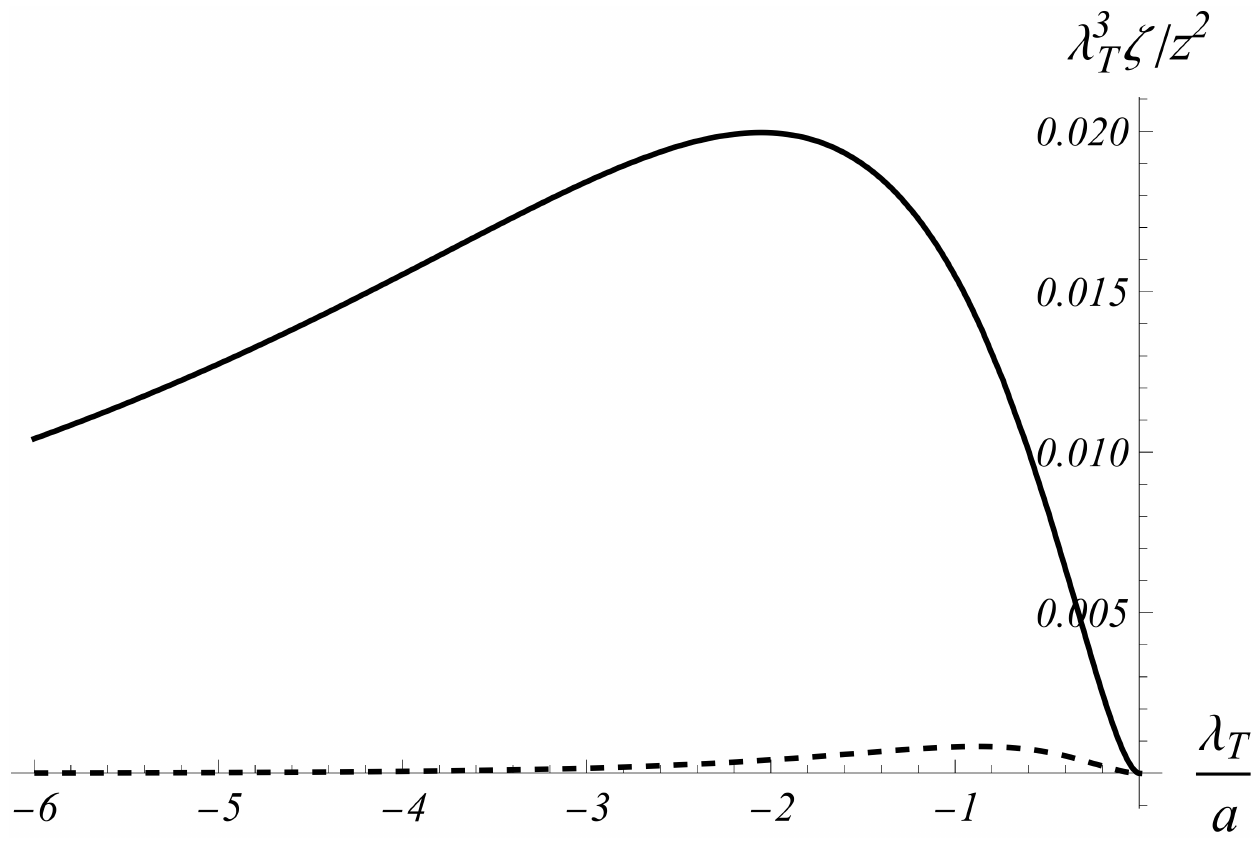}
  \end{center}
 \end{minipage}
&
 \begin{minipage}{0.47\hsize}
  \begin{center}
   \includegraphics[width=74mm]{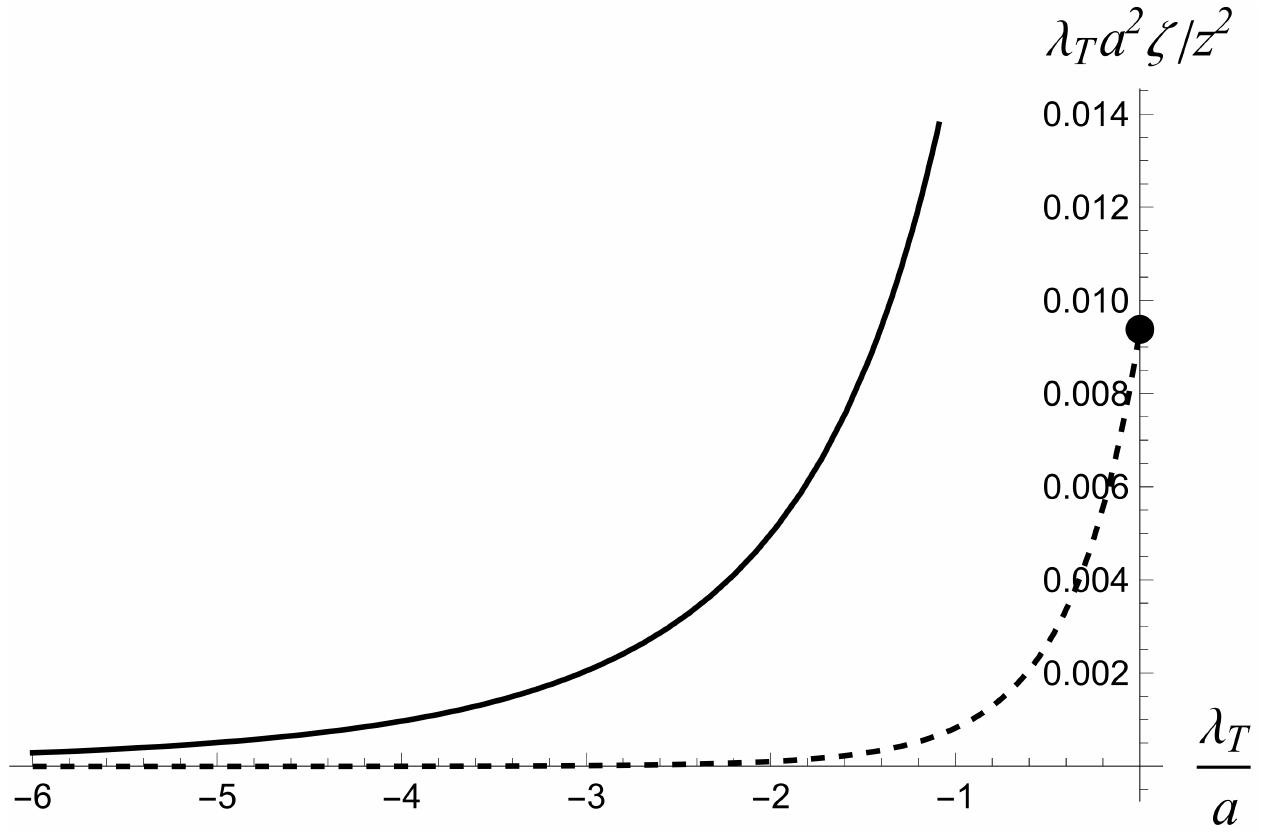}
  \end{center}
 \end{minipage}
\end{tabular}
\caption{Bulk viscosity in the high-temperature limit for $d=3$ as functions of $\lambda_{T}/a$ in thermal units as $\lambda_{T}^{3}\zeta/z^{2}$ (left) and in interaction units as $\lambda_{T}a^{2}\zeta/z^{2}$ (right).
The solid lines represent the total bulk viscosity $\zeta=\zeta_{\pair}+\zeta_{\fermion}$, which is much larger than the purely fermionic contribution $\zeta_{\fermion}$ (dashed lines).
The single data point in the right panel shows the fermionic contribution to
 the bulk viscosity at unitarity derived in Ref.~\cite{Dusling:2013},
 while the present work gives the complete bulk viscosity over the
 entire range of negative scattering lengths.\label{fig:bulk3D-negative}}
\end{figure}

Before we dive into the details, we give a brief summary of the results of this paper for the bulk viscosity.
We evaluate the bulk viscosity in the high-temperature regime, where the
fermion fugacity $z\equiv e^{\beta \mu}$ is small and serves as an expansion parameter with
$\beta$ the inverse temperature and $\mu$ the fermion chemical
potential, respectively~\cite{Liu:2013}.
We consider two-component fermions with a short-range interaction that is characterized by an $s$-wave scattering length $a$. At positive scattering length $a>0$ there exists a two-body bound state of two distinct fermions forming a bound pair. In this work, we focus on the fermion-dominated regime where the pair fugacity $z_{\text{pair}} = z^2e^{\beta/(ma^2)}$ is smaller than the fermion fugacity $z$. The parameter region in fugacity and interaction where this expansion applies is schematically shown in~Fig.~\ref{fig:expansion-regime}.
In the following we discuss the bulk viscosity separately for negative scattering lengths, when there is no two-body bound state, and for positive  scattering lengths when the bound state is present.
At negative scattering length in three dimensions the bulk viscosity $\zeta\sim z^2/\lambda_{T}^{3}$ is of second order in fugacity $z$ with the thermal de Broglie wavelength $\lambda_T\equiv\hbar\sqrt{2\pi \beta /m}$, as plotted in the left panel of~Fig.~\ref{fig:bulk3D-negative}.
We find that the bulk viscosity is divided into two contributions, $\zeta=\zeta_{\textrm{pair}}+\zeta_{\textrm{fermion}}$, where $\zeta_{\textrm{pair}}$ and $\zeta_{\textrm{fermion}}$ are computed as
\begin{equation}
\lambda_T^3\zeta_{\pair}
=z^2\frac{2\sqrt{2}}{9\pi}v^2
\Bigl[
    -1-(1+v^2)e^{v^2}\textrm{Ei}(-v^2)
\Bigr],
\xlabel{eq:result-bulk3Dpair-analitic-intro}
\end{equation}
and
\begin{equation}
\begin{split}
\lambda_{T}^{3}\zeta_{\fermion}
&= z^2\frac{v^2}{384\sqrt{2}}
    \frac{\left[2 \left(4+9 v^2 + 2 v^4\right)
 +\sqrt{\pi} e^{v^2} v \left(15+ 20v^2+4v^4\right) \text{erfc}(-v)\right]^2}{2- v^2+v^4+v^6 e^{v^2}\textrm{Ei}(-v^2)}\\
 &\quad\overset{v\to0}{\longrightarrow} \frac{z^2v^2}{12\sqrt2}\,,
\end{split}
\xlabel{eq:result-Fermi-pinch-bulk-RTA-intro}
\end{equation}
with dimensionless interaction parameter $v\equiv \lambda_T/(\sqrt{2\pi}a)=\hbar\sqrt{\beta/m}a^{-1}$.
 The pair contribution was computed previously from the contact correlation
function~\cite{Enss:2019,Nishida:2019,Hofmann:2020} without vertex corrections.
We find that vertex corrections contribute at the same order in fugacity due to
the pinch singularity and give rise to the additional fermionic contribution
$\zeta_{\fermion}$.  The fermion contribution at resonant interaction, $\lambda_{T}^{3}\zeta_{\fermion}(v\to0)=z^2v^2/12\sqrt2$,
agrees with a previous result from fermionic kinetic
theory~\cite{Dusling:2013}, but the extension to all scattering lengths
$a$ is new.
We show that both $\zeta_{\pair}$ and $\zeta_{\fermion}$ contributions individually are incomplete, and only together they yield the total bulk viscosity at leading order in fugacity.

On the other hand, at positive scattering length in both two and three dimensions, there are also bound states of pairs in addition to the scattering states of pairs included above. We show below that the bound pairs  exhibit their own version of the pinch singularity that arises for pair-pairhole excitations at vanishing frequency and momentum exchange. This pair pinch singularity has a dramatic effect on the pair bubble contribution to the bulk viscosity, which is reduced by one order in fugacity and thus contributes to the bulk viscosity already at first order in $z$ as $\zeta\sim ze^{\beta/ma^2}/\lambda_T^3$. This bound-pair term seems to dominate over the $O(z^2)$ contributions from scattering pairs and from fermions at high temperature. Furthermore, we find that the bound-pair vertex corrections are of the same order $O(z)$ due to the pair pinch singularity, and therefore the vertex corrections are equally important and need to be resummed. We present a self-consistent equation to resum these vertex corrections. While the numerical solution of this self-consistent equation is difficult and the scaling of the bound-pair contribution to the bulk viscosity remains an open question, the combination of the Boltzmann equation for fermions with the kinetic equation for bound pairs provides a consistent description of transport in strongly correlated many-body systems with bound states.

The paper is structured as follows:
In  Section~\ref{sec:microscopics}, we start with the general formulation of two-component fermions with a zero-range interaction and the Kubo formula for the bulk viscosity.
Next, we briefly review an order counting method with respect to the fugacity in the quantum virial expansion.
We then introduce two kinds of pinch singularities in the quantum virial expansion: the fermion pinch singularity and the pair pinch singularity.
The fermion (pair) singularity appears in the product of fermion (pair) propagators.
The fermion pinch singularity was introduced in Ref.~\cite{Fujii:2021} and
enhances the contribution of on-shell fermions, whereas the pair pinch
singularity is a new type of singularity and enhances the contribution of bound
pairs.
In  Section~\ref{sec:Resum-fermion-pinch}, we discuss the bulk viscosity for negative scattering lengths.
Since the system with a negative scattering length has no bound state and there is no pair pinch singularity, we only need to take into account the fermion pinch singularity.
We present the complete result of the bulk viscosity up to second order in fugacity.
In  Section~\ref{sec:Resum-pair-pinch}, we discuss the bulk viscosity for positive scattering lengths.
We show that the bulk viscosity appears at first order in fugacity due to the pinch singularity, and derive a self-consistent equation for a vertex function to compute the first-order bulk viscosity.
We also reduce the self-consistent equation to a kinetic equation for bound pairs by employing a simple approximation.
Finally,  Section~\ref{sec:summary} is devoted to a summary of this paper.

Our study partially follows the analysis described in Ref.~\cite{Fujii:2021}
and applies it to the bulk viscosity.
In what follows, we will set $\hbar=k_B=1$ and implicit sums over repeated spin indices $\sigma=\ \uparrow,\,\downarrow$ are assumed throughout this paper.
The bosonic and fermionic Matsubara frequencies are denoted by $\omega^{B}_{m}=2\pi m/\beta$ and $\omega^{F}_{m}=2\pi(m+1/2)/\beta$, respectively, for $m\in\mathbb{Z}$.
Also, an integration over $d$-dimensional wave vector or momentum is denoted by $\int_{\bp}\equiv\int\!d\bp/(2\pi)^d$ for the sake of brevity.

\section{Preliminary}

\xlabel{sec:microscopics}
\subsection{Microscopics}

Let us consider a two-component Fermi gas with a zero-range interaction in $d$ spatial dimensions, whose Hamiltonian is provided by
\begin{equation}
\hat{H}
= \int\!d\boldsymbol{x}\,
 \hat{\psi}^{\dagger}_{\sigma}(\bx)
 \left(
  -\frac{\nabla^{2}}{2m}
    \right)
 \hat{\psi}_{\sigma}(\bx)
+ \frac{g}{2}\int\!d\bx\,
 \hat{\psi}^{\dagger}_\sigma(\bx)
 \hat{\psi}^{\dagger}_{\sigma^{\prime}}(\bx)
 \hat{\psi}_{\sigma^{\prime}}(\bx)
 \hat{\psi}_{\sigma}(\bx).
\xlabel{eq:hamiltonian}
\end{equation}
We work in the Matsubara formalism at inverse temperature $\beta=1/T$ and chemical potential $\mu$, and write the bare fermion propagator in the Fourier space as
\begin{equation}
G(i\omega^{F}_{m},\bp) = \frac{1}{i\omega^{F}_{m}-\epsilon_{\bp}+\mu}
\xlabel{eq:fermion-bare-propag}
\end{equation}
and the full fermion propagator as
\begin{equation}
\calG(i\omega^{F}_{m},\bp) = \frac{1}{i\omega^{F}_{m}-\epsilon_{\bp}+\mu-\Sigma(i\omega^{F}_{m},\bp)},
\xlabel{eq:fermion-full-propag}
\end{equation}
where $\epsilon_{\bp}=\bp^2/(2~m)$ is the single-particle energy and $\Sigma(i\omega^{F}_{m},\bp)$ is the fermion self-energy.
We also introduce the pair propagator in vacuum, represented diagrammatically in  Fig.~\ref{fig:pair-propagator}, as
\begin{equation}
D(i\omega^{B}_{m},\bp)
= \frac{\Omega_{d-1}}{m} \frac{d-2}{a^{2-d}-[-m(i\omega^{B}_{m}-\epsilon_{\bp}/2+2\mu)]^{d/2-1}},
\xlabel{eq:pair-bare-propag}
\end{equation}
and the full pair propagator as
\begin{equation}
\calD(i\omega^{B}_{m},\bp)
= \frac{1}{D(i\omega^{B}_{m},\bp)^{-1}-\Delta(i\omega^{B}_{m},\bp)},
\xlabel{eq:pair-full-propag}
\end{equation}
where $\Omega_{d-1}\equiv(4\pi)^{d/2}/[2\Gamma(2-d/2)]=2,\,2\pi,\,4\pi$ coincides with the surface area of the unit $(d-1)$-sphere for $d=1,\,2,\,3$ and $\Delta(i\omega^{B}_{m},\bp) $ is the pair ``self-energy''.\footnote{Although $\Delta(i\omega^{B}_{m},\bp)$ does not have dimension of energy, we call it ``self-energy'' in analogy with the fermion self-energy.}
Here, the scattering length $a$ is related to the bare coupling $g$ in Eq.~\eqref{eq:hamiltonian} via
\begin{equation}
g = \frac{\Omega_{d-1}}{m}
\frac{d-2}{a^{2-d}-\Lambda^{d-2}/[\Gamma(d/2)\Gamma(2-d/2)]}
\end{equation}
in the cutoff regularization~\cite{Fujii:2020}.
We note that the bare pair propagator is simply the two-body scattering $T$-matrix.

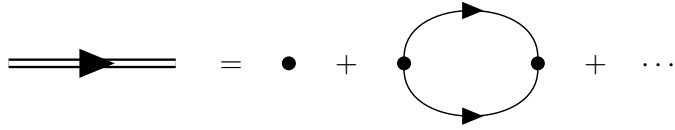
\begin{figure}[t]
\begin{equation*}
\scalebox{1.1}{
\begin{tikzpicture}[baseline=(o.base)]
\begin{feynhand}
	\vertex (o) at (0,-0.1);
	\vertex (a1) at (0,0); \vertex (a2) at (2,0);
	\propag [double,double distance=0.4ex,thick,with arrow=0.5,arrow size=1.0em] (a1) to (a2);
\end{feynhand}
\end{tikzpicture}
$\quad = \quad$
\begin{tikzpicture}[baseline=(o.base)]
\begin{feynhand}
	\vertex (o) at (0,-0.1);
	\vertex [dot] (a1) at (-1,0) {};
\end{feynhand}
\end{tikzpicture}
$\quad + \quad$
\begin{tikzpicture}[baseline=(o.base)]
\begin{feynhand}
	\vertex (o) at (0,-0.1);
	\vertex [dot] (a1) at (0,0) {}; \vertex [dot] (a2) at (1.6,0) {};
	\propag [fer,half left,looseness=1.2] (a1) to (a2);
	\propag [fer,half right,looseness=1.2] (a1) to (a2);
\end{feynhand}
\end{tikzpicture}
$\quad+\quad\cdots$
}
\end{equation*}
\caption{Diagrammatic representation of the bare pair propagator.
The dot denotes a bare coupling constant, while the thin single and double lines represent the bare fermion and pair propagators, respectively.\xlabel{fig:pair-propagator}}
\end{figure}

In the high-temperature regime, the fermion self-energy is evaluated as
\begin{equation}
\Sigma(i\omega^{F}_{m},\bp)
= z\int_{\bq} e^{-\beta\epsilon_{\bq}}
 D(i\omega^{F}_{m}+\epsilon_{\bq}-\mu,\bp+\bq)
+ O(z^2),
\xlabel{eq:fermion-self-energy}
\end{equation}
whose diagrammatic representation is depicted in the upper-left panel of  Fig.~\ref{fig:self-energy}.
Just as the fermion self-energy $\Sigma(i\omega^{F}_{m},\bp)$ is expressed with the two-body scattering $T$-matrix, the pair self-energy $\Delta(i\omega^{B}_{m},\bp)$ is expressed with the three-body (i.e.,~pair-fermion) scattering $T$-matrix $\calT_{3}$ as (See~\xref{app-sec:pair-self-energy} for the detailed derivation)
\begin{equation}
\begin{split}
\Delta(i\omega^{B}_{m},\bp)
&= 2z\int_{\bq} e^{-\beta\epsilon_{\bq}}
 [ G(i\omega^{B}_{m}-\epsilon_{\bq}+\mu,\bp+\bq) \\
&\quad 
 +\calT_{3}(i\omega^{B}_{m},\bp;\epsilon_{\bq}-\mu,\bq\vert i\omega^{B}_{m},\bp;\epsilon_{\bq}-\mu,\bq)]
+ O(z^2).
\end{split}\xlabel{eq:pair-self-energy}
\end{equation}
Note that we define $\Delta(i\omega^{B}_{m},\bp)$ by removing the $O(z^{0})$ part because it is included in the definition of $D(i\omega^{B}_{m},\bp)$, and thus $\Delta(i\omega^{B}_{m},\bp)$ is $O(z)$.
Here, $\calT_{3}(i\omega^{B}_{m},\bp;i\omega^{F}_{n},\bq\vert i\omega^{B}_{m^{\prime}},\bp^{\prime};i\omega^{F}_{n^{\prime}},\bq^{\prime})$ is the three-body scattering $T$-matrix in vacuum between a pair and a fermion from incoming $(i\omega^{B}_{m^\prime},\bp^\prime;i\omega^{F}_{n^\prime},\bq^\prime)$ to outgoing $(i\omega^{B}_{m},\bp;i\omega^{F}_{n},\bq)$ frequencies and momenta, and is expressed by the sum of an infinite number of diagrams depicted in the lower panel of  Fig.~\ref{fig:self-energy}.

\begin{figure}[t]
\begin{align*}
\scalebox{1.2}{
\begin{tikzpicture}[scale=1,transform shape,baseline=-0.5ex]
\begin{feynhand}
	\vertex (a1) at (0.2,0);
	\vertex [draw,circle] (a2) at (1.1,0) {$\Sigma$};
	\vertex (a3) at (2.0,0);
	\propag [fermion] (a1) to (a2);
	\propag [fermion] (a2) to (a3);
\end{feynhand}
\end{tikzpicture}
$\ =\ $
\begin{tikzpicture}[scale=1,transform shape,baseline=-0.5ex]
\begin{feynhand}
	\vertex (a1) at (0.6,0); \vertex (a2) at (1.2,0);
	\vertex (a3) at (2.4,0); \vertex (a4) at (3.0,0);
	\propag [fermion] (a1) to (a2);
	\propag [double,double distance=0.4ex,thick,with arrow=0.5,arrow size=0.8em] (a2) to (a3);
	\propag [fermion] (a3) to (a4);
	\propag [fermion, half right,looseness=1.5] (a3) to (a2);
\end{feynhand}
\end{tikzpicture}
$\qquad\quad$
\begin{tikzpicture}[scale=1,transform shape,baseline=-0.6ex]
\begin{feynhand}
	\vertex (b1) at (0,0);
	\vertex [draw,circle] (b2) at (1.0,0) {$\Delta$};
	\vertex (b3) at (2.0,0);
	\propag [double,double distance=0.4ex,thick,with arrow=0.5,arrow size=0.8em] (b1) to (b2);
	\propag [double,double distance=0.4ex,thick,with arrow=0.5,arrow size=0.8em] (b2) to (b3);
\end{feynhand}
\end{tikzpicture}
$\ =\ $
\begin{tikzpicture}[scale=1,transform shape,baseline=-0.3ex]
\begin{feynhand}
	\vertex (b1) at (0.5,0.045); \vertex (b2) at (1.2,0.045);
	\vertex (b3) at (1.9,0.045); \vertex (b4) at (2.6,0.045);
	\vertex (c1) at (1.2,0); \vertex (c2) at (1.9,0);
	\vertex (c3) at (1.2,0.7); \vertex (c4) at (1.9,0.7);
	\vertex [scale=1.1] (c0) at (1.55,0.32) {$\mathcal{T}_{3}$};
	\propag [thick] (c1) to (c2); \propag [thick] (c2) to (c4);
	\propag [thick] (c4) to (c3); \propag [thick] (c3) to (c1);
	\propag [double,double distance=0.4ex,thick,with arrow=0.5,arrow size=0.8em] (b1) to (b2);
	\propag [double,double distance=0.4ex,thick,with arrow=0.5,arrow size=0.8em] (b3) to (b4);
	\propag [fermion,half right,looseness=1.8] (c4) to (c3);
\end{feynhand}
\end{tikzpicture}
}
\end{align*} \\
\begin{align*}
\scalebox{1.1}{
\begin{tikzpicture}[scale=1,transform shape,baseline=2.4ex]
\begin{feynhand}
\vertex (b1) at (0.4,0.045); \vertex (b2) at (1.2,0.045);
\vertex (b3) at (2.2,0.045); \vertex (b4) at (3.0,0.045);
\vertex (c1) at (1.2,0); \vertex (c2) at (2.2,0);
\vertex (c3) at (1.2,1.05); \vertex (c4) at (2.2,1.05);
\vertex (a1) at (0.4,1.05); \vertex (a2) at (3.0,1.05);
\vertex [scale=1.2] (c0) at (1.7,0.5) {$\mathcal{T}_{3}$};
\propag [thick] (c1) to (c2); \propag [thick] (c2) to (c4);
\propag [thick] (c4) to (c3); \propag [thick] (c3) to (c1);
\propag [double,double distance=0.4ex,thick,with arrow=0.5,arrow size=0.8em] (b1) to (b2);
\propag [double,double distance=0.4ex,thick,with arrow=0.5,arrow size=0.8em] (b3) to (b4);
\propag [fermion] (a1) to (c3);
\propag [fermion] (c4) to (a2);
\end{feynhand}
\end{tikzpicture}
$\ = \ $
\begin{tikzpicture}[scale=1,transform shape,baseline=2.4ex]
\begin{feynhand}
\vertex (b1) at (0.4,0.045); \vertex (b2) at (1.2,0.045);
\vertex (a2) at (1.2,0); \vertex (c2) at (1.2,0.09);
\vertex (b3) at (1.8,1.0); \vertex (b4) at (2.6,1.0);
\vertex (a3) at (1.8,0.955); \vertex (c3) at (1.8,1.045);
\vertex (f1) at (0.4,1.045); \vertex (f2) at (2.6,0);
\propag [fermion] (f1) to (c3);
\propag [double,double distance=0.4ex,thick,with arrow=0.5,arrow size=0.8em] (b1) to (b2);
\propag [double,double distance=0.4ex,thick,with arrow=0.5,arrow size=0.8em] (b3) to (b4);
\propag [fermion] (a2) to (f2);
\propag [fermion] (c2) to (a3);
\end{feynhand}
\end{tikzpicture}
$+$
\begin{tikzpicture}[scale=1,transform shape,baseline=2.4ex]
\begin{feynhand}
\vertex (b1) at (0.4,0.045); \vertex (b2) at (1.2,0.045);
\vertex (a2) at (1.2,0); \vertex (c2) at (1.2,0.09);
\vertex (b3) at (1.8,1.0); \vertex (b4) at (2.6,1.0);
\vertex (a3) at (1.8,0.955); \vertex (c3) at (1.8,1.045);
\vertex (a4) at (2.6,0.955); \vertex (c4) at (2.6,1.045);
\vertex (b5) at (3.2,0.045); \vertex (b6) at (4.0,0.045);
\vertex (a5) at (3.2,0); \vertex (c5) at (3.2,0.09);
\vertex (f1) at (0.4,1.045); \vertex (f2) at (4.0,1.045);
\propag [double,double distance=0.4ex,thick,with arrow=0.5,arrow size=0.8em] (b1) to (b2);
\propag [double,double distance=0.4ex,thick,with arrow=0.5,arrow size=0.8em] (b3) to (b4);
\propag [double,double distance=0.4ex,thick,with arrow=0.5,arrow size=0.8em] (b5) to (b6);
\propag [fermion] (f1) to (c3);
\propag [fermion] (a2) to (a5);
\propag [fermion] (c4) to (f2);
\propag [fermion] (c2) to (a3);
\propag [fermion] (a4) to (c5);
\end{feynhand}
\end{tikzpicture}
$\ +\ \cdots$
}
\end{align*}
\caption{Diagrammatic representations of the fermion self-energy (upper left panel) and the pair self-energy (upper right panel) in Eqs.~\eqref{eq:fermion-self-energy} and \eqref{eq:pair-self-energy}. Here, $\calT_{3}$ is the three-body scattering $T$-matrix in vacuum between a pair and a fermion (lower panel).\xlabel{fig:self-energy}}
\end{figure}
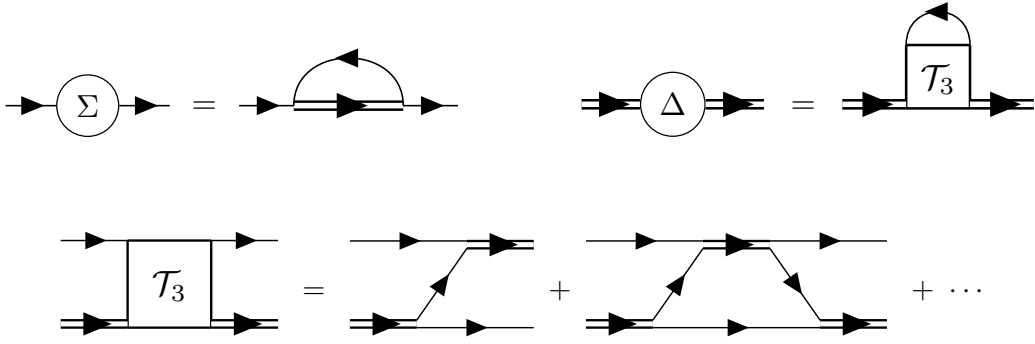

\subsection{Kubo formula}

According to linear response theory~\cite{Kubo:1957a,Kubo:1957b,Fujii:2020},
the Kubo formula for the bulk viscosity is provided by,
\begin{equation}
\zeta = \lim_{\omega \to 0}\frac{\Im[\chi_{\tilde{\Pi}}(\omega+i0^+)]}{\omega},
\xlabel{eq:Kubo-formula}
\end{equation}
where $\chi_{\calO}(\omega+i0^{+})$ is a retarded correlation function at zero momentum for an operator $\hat{\calO}$.
Here, $\hat{\tilde{\Pi}}$ is the modified trace of the stress tensor operator defined by
\begin{equation}
\hat{\tilde{\Pi}}
\equiv \frac{1}{d}\sum_{i=1}^{d}\hat{\Pi}_{ii}
 -\left(\pd{\calP}{\calN}\right)_{\calE,a}\hat{N}
 -\left(\pd{\calP}{\calE}\right)_{\calN,a}\hat{H},
\end{equation}
with $\hat{\Pi}_{ij}$ and $\hat{N}$ being the stress tensor and particle number operators and $\calP$, $\calN$, and $\calE$ the pressure, the particle number density, and the energy density, respectively.
In the Matsubara formalism, the retarded correlation function is obtained from the corresponding imaginary-time-ordered correlation function,
\begin{equation}
\xlabel{eq:correlation}
\chi_{\calO}(i\omega^{B}) = \frac{1}{L^d}\int^{\beta}_{0}\!d\tau\,e^{i\omega^{B}\tau}\average{\calT\,\hat{\calO}(\tau)\hat{\calO}(0)},
\end{equation}
with an analytic continuation of $i\omega^{B}\to\omega+i0^{+}$~\cite{Altland-Simons}.

In the system described by Eq.~\eqref{eq:hamiltonian}, the stress tensor
operator satisfies~\cite{Fujii:2018}
\begin{equation}
\sum_{i=1}^{d}\hat{\Pi}_{ii}
= 2\hat{H}+\frac{\hat{C}}{\Omega_{d-1}ma^{d-2}},
\end{equation}
which leads to
\begin{equation}
\hat{\tilde{\Pi}}
= \frac{\hat{C}}{d\Omega_{d-1}ma^{d-2}}
 -\left(\pd{\calP}{\calN}\right)_{\calE,a}\hat{N}
 -\left[\left(\pd{\calP}{\calE}\right)_{\calN,a}-\frac{2}{d}\right]\hat{H},
\xlabel{eq:modified-stress}
\end{equation}
with $\hat{C}$ being the contact operator defined by~\cite{Braaten:2008}
\begin{equation}
\hat{C}
\equiv\frac{(mg)^{2}}{2}\int\!d\bx\,
 \hat{\psi}^{\dagger}_\sigma(\bx)
 \hat{\psi}^{\dagger}_{\sigma^{\prime}}(\bx)
 \hat{\psi}_{\sigma^{\prime}}(\bx)
 \hat{\psi}_{\sigma}(\bx).
\xlabel{eq:def-contact-op}
\end{equation}
The expression~\eqref{eq:modified-stress} enables us to calculate the bulk viscosity via the contact correlation function, $\zeta = (d\Omega_{d-1}ma^{d-2})^{-2} \lim_{\omega\to 0} \Im[\chi_{C}(\omega+i0^{+})]/\omega + \cdots$, but one has to be careful that the other terms are non-negligible.\footnote{
Although the retarded correlation functions containing the number density and Hamiltonian operators vanish, when divided by $\omega$ they can give a finite contribution in the zero-frequency limit.
We see, in fact, that the terms proportional to the conserved quantities in Eq.~\eqref{eq:modified-stress} are essential in Eq.~\eqref{eq:Chapman-Enskog-condition}.
}
It is worth noting that the contact, which is a two-body operator, is essential for the bulk viscosity.
Because the correlation function of a one-body operator is dominant for the
shear viscosity and the thermal conductivity at high temperatures in contact
interacting systems~\cite{Enss:2011,Frank:2020,Fujii:2021}, the bulk viscosity
is inherently different from the other transport coefficients.\footnote{
In long-range interacting systems, one-body operator correlation functions do
not necessarily dominate for the shear viscosity and the thermal conductivity
at high temperatures~\cite{Nishida:2007,Link:2018,Galitski:2020}.
}

For later use, we note that the pressure fluctuations in Eq.~\eqref{eq:modified-stress} are calculated as
\begin{subequations}
\begin{align}
\left(\pd{\calP}{\calN}\right)_{\calE,a}
&= z\biggl[
 \frac{d+4}{d}\pd{b_{2}}{\beta}
 +\frac{2}{d} \beta\pd{\mbox{}^{2}b_{2}}{\beta^{2}}
\biggr]+O(z^{2}), \\
 \left(\pd{\calP}{\calE}\right)_{\calN,a}-\frac{2}{d}
&=- z \biggl[
 \frac{2}{d}\beta\pd{b_{2}}{\beta}
 +\frac{4}{d^{2}} \beta^{2} \pd{\mbox{}^{2}b_{2}}{\beta^{2}}
\biggr]+O(z^{2}),
\end{align}
\xlabel{eq:pressure-fluct}
\end{subequations}
with the virial expansion of the pressure $\calP=\frac{2}{\beta\lambda_{T}^{d}}[z+b_{2}(\beta, a)z^{2}+O(z^{3})]$.
The second virial coefficient $b_{2}(\beta,a)$ is given by~\cite{Nishida:2019}
\begin{equation}
b_{2}(\beta,a)
= -\frac{1}{2^{d/2+1}}
 -\frac{2^{d/2-2}}{\Omega_{d-1}}
 \int^{\infty}_{-\infty}\frac{d\varepsilon}{\pi}\,
 e^{-\beta\varepsilon}\Im\left[
  \frac{m^{2}D(\varepsilon-2\mu+i0^{+},\bzero)}
   {(-m\varepsilon-i0^{+})^{2-d/2}}
 \right].
\xlabel{eq:second-virial-coeff}
\end{equation}
With the use of the binding energy $\epsilon_B\equiv 1/(ma^2)$ and
\begin{equation}
\begin{split}
&\Im\biggl[
 \frac{m^{2}D(\varepsilon-2\mu-i0^{+},\bzero)}{(-m\varepsilon+i0^{+})^{2-d/2}}
\biggr]
\\
&\quad= 
\begin{cases}
\dis
4\pi^{2}\delta(\varepsilon+\epsilon_{B})
-\theta(\varepsilon)
 \frac{m^{3}}{4}
 \frac{| D(\varepsilon-2\mu-i0^{+},\bzero)| ^{2}}{m\varepsilon}
& (d=2), \\
\dis 
\theta(a)8\pi^{2}\delta(\varepsilon+\epsilon_{B})
-\theta(\varepsilon)
 \frac{m^{3}}{4\pi a}
 \frac{| D(\varepsilon-2\mu-i0^{+},\bzero)| ^{2}}{\sqrt{m\varepsilon}}
& (d=3),
\end{cases}
\end{split}\xlabel{eq:integrand-of-b2}
\end{equation}
the second virial coefficient is divided into the bound- and scattering-state contributions as
\begin{equation}
b_2(\beta,a)
= -\frac{1}{2^{d/2+1}}
+ \delta b^{(\textrm{bound})}_{2}(\beta,a)
+ \delta b^{(\textrm{scat})}_{2}(\beta,a),
\xlabel{eq:b2virial}
\end{equation}
with
\begin{equation}
\delta b^{(\textrm{bound})}_{2}(\beta,a)
=
\begin{cases}
\dis e^{\beta \epsilon_B} & (d=2), \\
\dis \theta(a)\sqrt{2}e^{\beta \epsilon_B} & (d=3),
\end{cases}\end{equation}
and
\begin{equation}
\delta b^{(\textrm{scat})}_{2}(\beta,a)
=
\begin{cases}
\dis -\int^{\infty}_{0}\frac{d\varepsilon}{\varepsilon}\frac{e^{-\beta\varepsilon}}{[\ln(ma^2\varepsilon)]^2+\pi^2} & (d=2), \\
\dis -\frac{1}{\sqrt{2m}\pi a}\int^{\infty}_{0}d\varepsilon\,\frac{e^{-\beta\varepsilon}}{\sqrt{\varepsilon}(\varepsilon+\epsilon_B)} & (d=3).
\end{cases}\end{equation}
These expressions separating the bound- and scattering-state contributions are
consistent with the standard Beth--Uhlenbeck expression for the second virial
coefficient~\cite{Beth:1937} and were derived for $d=2$ in
Refs.~\cite{Chafin:2013,Ngampruetikorn:2013,Barth:2014} and for $d=3$ in
Ref.~\cite{Leyronas:2011}.\footnote{
In three dimensions, the second virial coefficient can also be expressed simply
as $b_2(\beta,a)=-\frac{1}{4\sqrt{2}}+\frac{1}{\sqrt{2}}\bigl(1+\textrm{erf}(\sqrt{\beta/m}a^{-1})\bigr)e^{\beta \epsilon_B}$, which is a monotonic function of the inverse scattering
length~\cite{Bedaque:2003,Enss:2019}.
}

\subsection{Order counting in the quantum virial expansion}

\xlabel{sec:order-counting-method}
We compute the bulk viscosity via the imaginary-time-ordered correlation function in the quantum virial expansion.
When we use the propagators in the Matsubara frequency representation, the fugacity appears through the expansion of the distribution function resulting from the Matsubara frequency summation.
It is a little tricky to find the fugacity dependence in the Matsubara frequency representation before summing over the Matsubara frequency.
For this reason, it is helpful to use the knowledge of the propagators in the imaginary time representation only when we count the order of diagrams.
The bare fermion propagator in the imaginary-time representation has an explicit fugacity dependence,
\begin{equation}
G(\tau,\bp)
= -e^{-\tau(\epsilon_{\bp}-\mu)}[\theta(\tau)-f_{F}(\epsilon_{\bk}-\mu)]
= \sum_{n=0}^{\infty}z^{n}G^{(n)}(\tau,\bp)
\end{equation}
with
\begin{equation}
G^{(n)}(\tau,\bp)
=
\begin{cases}
\dis -e^{-\tau(\epsilon_{\bp}-\mu)}\theta(\tau) & (n=0),\\
\dis -e^{-\tau(\epsilon_{\bp}-\mu)}\bigl(-e^{\beta\epsilon_{\bp}}\bigr)^{n} & (n\geq 1), 
\end{cases}\end{equation}
where $f_{F}(\varepsilon)=1/(e^{\beta\varepsilon}-1)$ is the Fermi distribution function and $\theta(\tau)$ the step function.
Using $G^{(n)}(\tau,\bk)$, we can directly provide the coefficients of each order for the fugacity in the quantum virial expansion.
In fact, the diagrammatic computational method for the quantum virial expansion
is established in the imaginary-time representation, not in the Matsubara
frequency representation~\cite{Kaplan:2011,Leyronas:2011}.
In particular, the zeroth-order component $G^{(0)}(\tau,\bp)$ runs only in the forward direction for the imaginary time because it involves $\theta(\tau)$.
Thus, we can estimate the order of each diagram in the fugacity from the number of propagators going backward in imaginary time when an imaginary time is assigned to each vertex.
In fact, each of the self-energy diagrams in the upper panel of  Fig.~\ref{fig:self-energy} has just one propagator running backward in imaginary time (running from right to left).
Nevertheless, we mainly employ the Matsubara frequency representation because it is more convenient than the imaginary-time representation for the resummation discussed later.

\subsection{Pinch singularity}

In the zero-frequency limit of the Kubo formula, there emerges the product of the retarded and advanced propagators with the same frequency and momentum.
For the fermion propagator, the product is decomposed into partial fractions as
\begin{equation}
\calG^{R}(\varepsilon,\bp)\calG^{A}(\varepsilon,\bp)
=\frac{\Im[\calG^{R}(\varepsilon,\bp)]}{\Im[\Sigma(\varepsilon+i0^{+},\bp)]},
\end{equation}
with $\calG^{R}(\varepsilon,\bp)=\calG(\varepsilon+i0^{+},\bp)$ and $\calG^{A}(\varepsilon,\bp)=\calG(\varepsilon-i0^{+},\bp)$.
Because of $\Im[\calG^{R}(\varepsilon,\bp)]=-\pi\delta(\varepsilon-\epsilon_{\bp}+\mu)+O(z)$ and $\Im[\Sigma(\varepsilon+i0^{+},\bp)]\sim O(z)$ in the high-temperature regime, the product is expressed as
\begin{equation}
\calG^{R}(\varepsilon,\bp)\calG^{A}(\varepsilon,\bp)
=\frac{\pi \delta(\varepsilon-\epsilon_{\bp}+\mu)}{-\Im[\Sigma(\varepsilon+i0^{+},\bp)]}+O(z^{0}),
\xlabel{eq:fermion-pinch}
\end{equation}
and is proportional to the inverse of the fugacity.
Thus, the appearance of the product \eqref{eq:fermion-pinch} complicates the
order counting in fugacity; this is the so-called pinch
singularity~\cite{Eliashberg:1962,Jeon:1995,Jeon:1996,Hidaka:2011}.
The order counting discussed in the previous section is valid only at nonzero frequencies and is not valid in the zero-frequency limit due to the pinch singularity.

We now move on to discuss the pinch singularity of the pair propagator.
In contrast to the bare fermion propagator $G^{R}(\varepsilon,\bp)$, the inverse of the bare pair propagator $D^{R}(\varepsilon,\bp)$ has a finite imaginary part,
\begin{equation}
\Im[D^{R}(\varepsilon,\bp)^{-1}]
= \theta(\varepsilon-\epsilon_{\bp}/2+2\mu)\times
\begin{cases}
\dis \frac{m}{4} & (d=2), \\
\dis \frac{m}{4\pi}\sqrt{m(\varepsilon-\epsilon_{\bp}/2+2\mu)} & (d=3),
\end{cases}\end{equation}
so that the product is decomposed into partial fractions as
\begin{equation}
\calD^{R}(\varepsilon,\bp)\calD^{A}(\varepsilon,\bp)
= \frac{\Im[\calD^{R}(\varepsilon,\bp)]}
 {
  \Im[\Delta(\varepsilon+i0^{+},\bp)]
  -\Im[D^{R}(\varepsilon,\bp)^{-1}]
 },
\end{equation}
with $\calD^{R}(\varepsilon,\bp)=\calD(\varepsilon+i0^{+},\bp)$ and $\calD^{A}(\varepsilon,\bp)=\calD(\varepsilon-i0^{+},\bp)$.
The imaginary part of the pair propagator is provided by
\begin{equation}
\begin{split}
\Im[\calD^{R}(\varepsilon,\bp)]
&= -\theta (a)\frac{2\pi \Omega_{d-1}}{m^{2}a^{4-d}}
 \delta(\varepsilon-\epsilon_{\bp}/2+2\mu+\epsilon_{B})
\\
&\quad - \theta(\varepsilon-\epsilon_{\bp}/2+2\mu)
\rho_{D}(\varepsilon-\epsilon_{\bp}/2+2\mu)+O(z),
\end{split}\xlabel{eq:pair-spectrum}
\end{equation}
where $\rho_{D}(\varepsilon)$ is the scattering continuum of $D^{R}(\varepsilon,\bp)$ defined by
\begin{equation}
\rho_{D}(\varepsilon)
\equiv-\Im[D^{R}(\varepsilon-2\mu,\bzero)]\vert _{\varepsilon>0}
=
\begin{cases}
\dis \frac{4\pi^{2}}{m}\frac{1}{[\ln(ma^{2}\varepsilon)]^{2}+\pi^{2}} & (d=2),\\
\dis \frac{4\pi}{m}\frac{\sqrt{m\varepsilon}}{a^{-2}+m\varepsilon} & (d=3).
\end{cases}\xlabel{eq:scattering-continuum}
\end{equation}
Therefore, together with $\Im[\Delta(\varepsilon+i0^{+},\bp)]\sim O(z)$ we find that
\begin{equation}
\calD^{R}(\varepsilon,\bp)\calD^{A}(\varepsilon,\bp)
= \theta (a)\frac{2\pi \Omega_{d-1}}{m^{2}a^{4-d}}
\frac{\delta(\varepsilon-\epsilon_{\bp}/2+2\mu+\epsilon_{B})}
 {
  -\Im[\Delta(\varepsilon+i0^{+},\bp)]
 }+O(z^{0})
 \xlabel{eq:pair-pinch}
\end{equation}
is also proportional to the inverse of the fugacity.
This singularity appears only when the scattering length is positive and enhances the contribution of the bound pair whose energy is given by the sum of the kinetic energy $\epsilon_{\bp}/2-2\mu$ and the binding energy $-\epsilon_{B}$.
Our goal is to give the exact bulk viscosity taking into account the fermion and pair pinch singularities, Eqs.~\eqref{eq:fermion-pinch} and \eqref{eq:pair-pinch}, respectively.

\section{Resummation of the fermion pinch singularity}

\xlabel{sec:Resum-fermion-pinch}
Let us first consider the case where only the fermion pinch singularity \eqref{eq:fermion-pinch} is relevant, which corresponds to systems with a negative scattering length in three dimensions.
For later convenience, we proceed with the discussion keeping it in general $d$ dimensions.

\subsection{Contact correlation function}

We first consider the contact correlation function to find the fermion pinch singularity.
If the expansion with respect to the fugacity at nonzero frequencies is applied, its leading term at $O(z^{2})$ is simply provided from a pair-bubble diagram, depicted in the first term of  Fig.~\ref{fig:contact-correlation}, whose contribution is expressed as
\begin{equation}
\chi_{C;\pair}(i\omega^{B})
= \frac{m^{4}}{\beta} \sum_{m}\int_{\bp}
 D(i\omega^{B}_{m}+i\omega^{B},\bp)
 D(i\omega^{B}_{m},\bp)+O(z^{3}).
\end{equation}
Here, it is sufficient to evaluate the correlation function with the full pair propagator replaced by the bare one for the leading order in fugacity.
The pair-bubble contribution has no fermion pinch singularity because it has no product of the fermion propagators.
With the use of the spectral representation of the pair propagator,
\begin{equation}
D(w,\bp)
=\int^{\infty}_{-\infty}\frac{d\varepsilon}{\pi}
 \frac{\Im[D(\varepsilon-i0^{+},\bp)]}{w-\varepsilon},
\end{equation}
the pair-bubble contribution is calculated as~\cite{Fujii:2020}
\begin{equation}
\begin{split}
\chi_{C;\pair}(i\omega^{B})
& = -z^{2}\frac{2^{d/2}m^{4}}{\lambda^{d}_{T}}
 \int^{\infty}_{-\infty}\frac{d\varepsilon}{\pi}
 \int^{\infty}_{-\infty}\frac{d\varepsilon^{\prime}}{\pi}
 \frac{e^{-\beta\varepsilon}-e^{-\beta\varepsilon^{\prime}}}
  {i\omega^{B}+\varepsilon-\varepsilon^{\prime}} \\
&\quad\times
 \Im[D(\varepsilon-2\mu-i0^{+},\bzero)]
 \Im[D(\varepsilon^{\prime}-2\mu-i0^{+},\bzero)]+O(z^{3}).
\end{split}\xlabel{eq:pair-bubble-correlation}
\end{equation}

\begin{figure}[t]
\begin{align*}
\scalebox{1.2}{
\begin{tikzpicture}[scale=1,transform shape,baseline=-0.6ex]
\begin{feynhand}
	\vertex [crossdot] (a1) at (0,0) {}; \vertex [crossdot] (a2) at (2,0) {};
	\propag [double,double distance=0.4ex,thick,with arrow=0.5,arrow size=1.0em,half right,looseness=1.0] (a1) to (a2);
	\propag [double,double distance=0.4ex,thick,with arrow=0.5,arrow size=1.0em,half right,looseness=1.0] (a2) to (a1);
\end{feynhand}
\end{tikzpicture}
$\ + \ $
\begin{tikzpicture}[scale=1,transform shape,baseline=-0.6ex]
\begin{feynhand}
	\vertex [crossdot] (a1) at (0,0) {};
	\vertex (b1) at (1.0,-0.8); \vertex (b2) at (2.0,-0.8);
	\vertex (b3) at (1.0,0.8); \vertex (b4) at (2.0,0.8);
	\vertex (c1) at (1.0,-0.845); \vertex (c2) at (2.0,-0.845);
	\vertex (c3) at (1.0,0.845); \vertex (c4) at (2.0,0.845);
	\vertex [crossdot] (a2) at (3.0,0) {};
	\propag [double,double distance=0.4ex,thick,with arrow=0.5,arrow size=0.8em,quarter right,looseness=0.5] (b3) to (a1);
	\propag [double,double distance=0.4ex,thick,with arrow=0.5,arrow size=0.8em,quarter right,looseness=0.5] (a1) to (b1);
	\propag [double,double distance=0.4ex,thick,with arrow=0.5,arrow size=0.8em,quarter right,looseness=0.5] (b2) to (a2);
	\propag [double,double distance=0.4ex,thick,with arrow=0.5,arrow size=0.8em,quarter right,looseness=0.5] (a2) to (b4);
	\propag [fermion] (c1) to (c2);
	\propag [fermion] (c1) to (c3);
	\propag [fermion] (c4) to (c3);
	\propag [fermion] (c4) to (c2);
\end{feynhand}
\end{tikzpicture}
$\ +\ \cdots$
}
\end{align*}
\caption{Diagrammatic representation of the imaginary-time-ordered contact correlation function.
The crossdot represents the bare vertex for the contact density (see the caption of  Fig.~\ref{fig:pair-propagator} for further notation).
We refer to the first term as the pair-bubble diagram and the second term as the box diagram.
Only the pair-bubble diagram provides the lowest-order contribution at $O(z^{2})$ at nonzero frequencies.\xlabel{fig:contact-correlation}}
\end{figure}
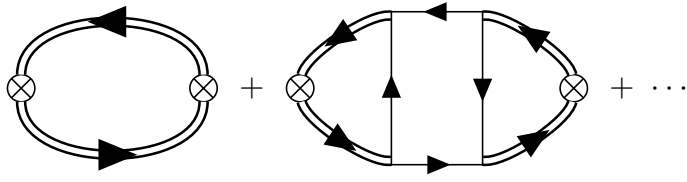

The next-to-leading-order contribution at $O(z^3)$ at nonzero frequencies is provided from an infinite set of diagrams.
Among them, the diagram with one fermion pinch singularity becomes $O(z^{2})$ in the zero-frequency limit and has a comparable contribution to $\chi_{C;\pair}(i\omega^{B})$.
For diagrams to have a product of the fermion propagators with the same frequency and momentum in the zero-frequency limit, they must have the shape where the left and right sides are connected just by the two fermion propagators, which we refer to as the fermion-bubble shape.
Among the infinite series of diagrams with $O(z^{3})$, only the second term in  Fig.~\ref{fig:contact-correlation}, which we call the box diagram, has such a fermion-bubble shape. 
The box-diagram contribution at nonzero frequencies is given by
\begin{equation}
\begin{split}
\chi_{C;\rmbox}(i\omega^{B})
&= \frac{2}{\beta} \sum_{m}\int_{\bp}
 C^{F}(i\omega^{F}_{m}+i\omega^{B},i\omega^{F}_{m};\bp)
  G(i\omega^{F}_{m}+i\omega^{B},\bp)\\
 &\quad \times G(i\omega^{F}_{m},\bp)
 C^{F}(i\omega^{F}_{m}+i\omega^{B},i\omega^{F}_{m};\bp),
\end{split}\xlabel{eq:box-diagram}
\end{equation}
with
\begin{equation}
C^{F}(i\omega^{F}_{m}+i\omega^{B},i\omega^{F}_{m};\bp)
 \equiv \frac{m^{2}}{\beta}\sum_{n}\int_{\bq}
 D(i\omega^{B}_{n}+i\omega^{B},\bq)
 D(i\omega^{B}_{n},\bq)
 G(i\omega^{B}_{n}-i\omega^{F}_{m},\bq-\bp).
\xlabel{eq:tilde-V}
\end{equation}
We can regard Eq.~\eqref{eq:box-diagram} as the fermion-bubble shape; the two vertex functions $C^{F}(i\omega^{F}_{m}+i\omega^{B},i\omega^{F}_{m};\bp)$ are connected by the two fermion propagators.
The product of the fermion propagators in Eq.~\eqref{eq:box-diagram} replaced by the full propagators results in the fermion pinch singularity, so that the box-diagram contribution becomes $O(z^{2})$ in the zero-frequency limit.
Therefore, the box-diagram contributions need to be resummed.
Since the fermion-bubble shape is the same as that of the diagrams providing
the shear viscosity and the thermal conductivity at high
temperatures~\cite{Fujii:2021}, the box diagram allows us to find other
contributions which have to be resummed.
They are provided as diagrams in which the four-point functions for the Maki--Thompson and Aslamazov--Larkin types and their iterations are inserted between the vertex functions in the box diagram.

\subsection{Pair-bubble and fermion pinch-singular contributions}

\xlabel{sec:pair-bubble-and-fermion-pinch}
To combine $C^{F}(i\omega^{F}_{m}+i\omega^{B},i\omega^{F}_{m};\bp)$ with the second and third terms of Eq.~\eqref{eq:modified-stress}, we introduce the vertex function $\tilde{\Pi}^{F}(i\omega^{F}_{m}+i\omega^{B},i\omega^{F}_{m};\bp)$ as
\begin{equation}
\tilde{\Pi}^{F}(i\omega^{F}_{m}+i\omega^{B},i\omega^{F}_{m};\bp)
\equiv 
 \frac{C^{F}(i\omega^{F}_{m}+i\omega^{B},i\omega^{F}_{m};\bp)}
  {d\Omega_{d-1}ma^{d-2}}
 -\left(\pd{\calP}{\calN}\right)_{\calE,a}
 -\left[\left(\pd{\calP}{\calE}\right)_{\calN,a}-\frac{2}{d}\right]\epsilon_{\bp}.
\xlabel{eq:fermion-vertex}
\end{equation}
The first term is evaluated as
\begin{equation}
\begin{split}
&\frac{C^{F}(i\omega^{F}_{m}+i\omega^{B},i\omega^{F}_{m};\bp)}{d\Omega_{d-1}m a^{d-2}} \\
& = \frac{zm}{d\Omega_{d-1}a^{d-2}} \int_{\bq}
 e^{-\beta\epsilon_{\bq}}
 D(\epsilon_{\bq}-\mu+i\omega^{F}_{m}+i\omega^{B},\bp+\bq)
 D(\epsilon_{\bq}-\mu+i\omega^{F}_{m},\bp+\bq)
 + O(z^{2}),
\end{split}\xlabel{eq:tilde-V-z1}
\end{equation}
so that $\tilde{\Pi}^{F}(i\omega^{F}_{m}+i\omega^{B},i\omega^{F}_{m};\bp)$ together with Eq.~\eqref{eq:pressure-fluct}
 is $O(z)$.
Thus, $\tilde{\Pi}^{F}(i\omega^{F}_{m}+i\omega^{B},i\omega^{F}_{m};\bp)$ can be interpreted as an effective vertex function for $\tilde{\Pi}$ at $O(z)$ connecting to the two fermionic propagators.

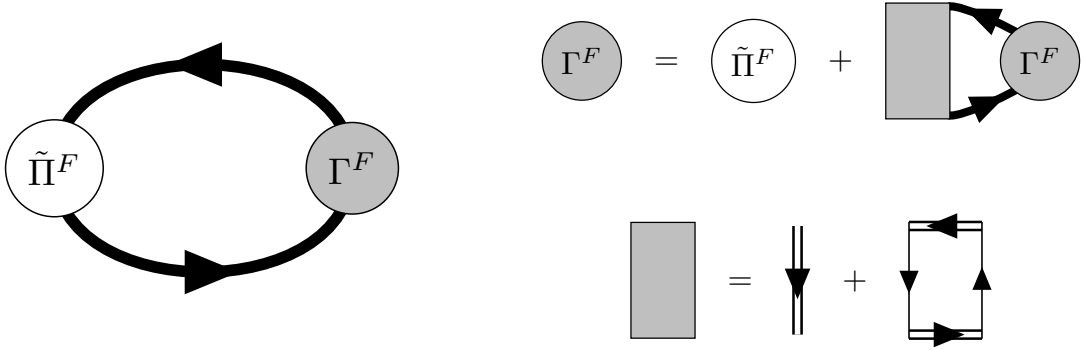
\begin{figure}[t]
\begin{tabular}{cc}
\begin{minipage}{0.36\hsize}
\begin{align*}
\scalebox{1.4}{
\begin{tikzpicture}
\begin{feynhand}
	\vertex (b1) at (0,0);
	\vertex (b2) at (2.8,0);
	\propag [line width=1.1mm,half right,looseness=1.2,with arrow=0.5,arrow size=1.2em] (b1) to (b2);
	\propag [line width=1.1mm,half right,looseness=1.2,with arrow=0.5,arrow size=1.2em] (b2) to (b1);
	\vertex [draw,fill=white,circle] (a1) at (0,0) {$\tilde{\Pi}^{F}$};
	\node [draw,fill=lightgray,circle] (c2) at (2.8,0) {$\Gamma^{F}$};
\end{feynhand}
\end{tikzpicture}
}
\end{align*}
\end{minipage}
\begin{minipage}{0.6\hsize}
\begin{align*}
\scalebox{1.2}{
\begin{tikzpicture}[scale=1,baseline=-0.6ex]
\begin{feynhand}
	\vertex [draw,fill=lightgray,circle] (a2) at (2.8,0) {$\Gamma^{F}$};
\end{feynhand}
\end{tikzpicture}
$\ \ =\ \ $
\begin{tikzpicture}[scale=1,baseline=-0.6ex]
\begin{feynhand}
	\vertex [draw,fill=white,circle] (a1) at (0,0) {$\tilde{\Pi}^{F}$};
\end{feynhand}
\end{tikzpicture}
$\ \ +\  \ $
\begin{tikzpicture}[scale=1,baseline=-0.6ex]
\begin{feynhand}
	\vertex (b2) at (0.8,-0.6);
	\vertex (b4) at (0.8,0.6);
	\vertex [circle] (a5) at (1.78,0) {};
	\propag [line width=0.9mm, quarter right, looseness=0.3, with arrow=0.4, arrow size=0.9em] (b2) to (a5);
	\propag [line width=0.9mm, quarter right, looseness=0.3, with arrow=0.6, arrow size=0.9em] (a5) to (b4);
	\vertex [draw,fill=lightgray,scale=3,yscale=1.82] (c1) at (0.46,0) {};
	\vertex [draw,fill=lightgray,circle] (a1) at (1.8,0) {$\Gamma^{F}$};
\end{feynhand}
\end{tikzpicture}
}
\end{align*}
\\
\begin{align*}
\scalebox{1.2}{
\begin{tikzpicture}[scale=1,baseline=-0.6ex]
\begin{feynhand}
	\vertex [draw,fill=lightgray,scale=3,yscale=1.82] (c1) at (0.46,0) {};
\end{feynhand}
\end{tikzpicture}
$\ \ =\ \ $
\begin{tikzpicture}[scale=1,baseline=-0.6ex]
\begin{feynhand}
	\vertex (a1) at (0,-0.6);
	\vertex (a2) at (0,0.6);
	\propag [double,double distance=0.4ex,thick,with arrow=0.5,arrow size=0.8em] (a2) to (a1);
\end{feynhand}
\end{tikzpicture}
$\ \ +\ \ $
\begin{tikzpicture}[scale=1,baseline=-0.6ex]
\begin{feynhand}
	\vertex (a1) at (0,-0.6);
	\vertex (a2) at (0.8,-0.6);
	\vertex (a3) at (0,0.6);
	\vertex (a4) at (0.8,0.6);
	\vertex (b1) at (0,-0.645);
	\vertex (b2) at (0.8,-0.645);
	\vertex (b3) at (0,0.645);
	\vertex (b4) at (0.8,0.645);
	\propag [double,double distance=0.4ex,thick,with arrow=0.5,arrow size=0.8em] (a4) to (a3);
	\propag [double,double distance=0.4ex,thick,with arrow=0.5,arrow size=0.8em] (a1) to (a2);
	\propag [fermion] (b3) to (b1);
	\propag [fermion] (b2) to (b4);
\end{feynhand}
\end{tikzpicture}
}
\end{align*}
\end{minipage}
\end{tabular}
\caption{(Left) Diagrammatic representation of Eq.~\eqref{eq:fermion-bubble-Pi-correlation}, which is the imaginary-time-ordered correlation function for $\tilde{\Pi}$ with the fermion pinch singularity.
(Upper right) Diagrammatic representation of the self-consistent equation for the vertex function $\Gamma^{F}$ to resum all relevant contributions.
(Lower right) Maki--Thompson and Aslamazov--Larkin diagrams for the four-point function represented by the rectangle.
The thick line represents the full fermion propagator, whereas the white and gray bulbs denote the bare vertex function $\tilde{\Pi}^{F}$ and the full one $\Gamma^{F}$, respectively.
It is sufficient for the leading-order vertex function to evaluate the four-point functions with the full propagators replaced by the free ones.\xlabel{fig:fermion-bubble}}
\end{figure}

With the use of $\tilde{\Pi}^{F}(i\omega^{F}_{m}+i\omega^{B},i\omega^{F}_{m};\bp)$, we can write down the correlation function for $\hat{\tilde{\Pi}}$ with the fermion pinch singularity.
To resum all the relevant contributions, we start with the formal expression of the correlation function for $\hat{\tilde{\Pi}}$ as
\begin{equation}
\begin{split}
\chi_{\tilde{\Pi};\fermion}(i\omega^{B})
&\equiv \frac{2}{\beta} \sum_{m}\int_{\bp}
    \tilde{\Pi}^{F}(i\omega^{F}_{m}+i\omega^{B},i\omega^{F}_{m};\bp)
 \calG(i\omega^{F}_{m}+i\omega^{B},\bp)\\
 &\quad\times\calG(i\omega^{F}_{m},\bp)
 \Gamma^{F}(i\omega^{F}_{m}+i\omega^{B},i\omega^{F}_{m};\bp),
\end{split}\xlabel{eq:fermion-bubble-Pi-correlation}
\end{equation}
which includes all contributions with the fermion pinch singularity.
The diagrammatic representation of Eq.~\eqref{eq:fermion-bubble-Pi-correlation} is depicted in the left panel of  Fig.~\ref{fig:fermion-bubble}.
The spin degeneracy is accounted for by the prefactor of $2$ and $\Gamma^{F}(i\omega^{F}_{m}+i\omega^{B},i\omega^{F}_{m};\bp)$ represents the full vertex function.

The correlation function for $\hat{\tilde{\Pi}}$ up to $O(z^{2})$ taking into account the fermion pinch singularity are fully incorporated in
\begin{equation}
\chi_{\tilde{\Pi}}(i\omega^{B})
= \frac{\chi_{C;\pair}(i\omega^{B})}{(d\Omega_{d-1}ma^{d-2})^{2}}
 + \chi_{\tilde{\Pi};\fermion}(i\omega^{B})
 + O(z^{3}).
\end{equation}
Accordingly, the bulk viscosity up to $O(z^{2})$ is obtained as
\begin{equation}
\zeta=\zeta_{\pair}+\zeta_{\fermion}+O(z^{3})
\xlabel{eq:complete-bulk-z2}
\end{equation}
with
\begin{align}
\zeta_{\pair}
&\equiv\lim_{\omega\to 0}
 \frac{1}{(d\Omega_{d-1}ma^{d-2})^{2}}
 \frac{\Im[\chi_{C;\pair}(\omega+i0^{+})]}{\omega},
 \xlabel{eq:PB-bulk} \\
\zeta_{\fermion}
&\equiv\lim_{\omega\to 0}
 \frac{\Im[\chi_{\tilde{\Pi};\fermion}(\omega+i0^{+})]}{\omega}.
 \xlabel{eq:fermion-pinch-bulk}
\end{align}
Eq.~\eqref{eq:complete-bulk-z2} provides the complete bulk viscosity up to $O(z^{2})$ as far as the pair pinch singularity does not occur.
The substitution of Eq.~\eqref{eq:pair-bubble-correlation} into~\eqref{eq:PB-bulk} leads to
\begin{equation}
\lambda^{d}_{T}\zeta_{\pair}
 = z^{2}\frac{2^{d/2}\beta m^{2}}{(d\Omega_{d-1}a^{d-2})^{2}}
 \int^{\infty}_{0}\frac{d\varepsilon}{\pi}
 e^{-\beta\varepsilon}
 \rho_{D}(\varepsilon)^{2},
\xlabel{eq:result-PB-bulk}
\end{equation}
which was obtained in the previous
studies~\cite{Nishida:2019,Enss:2019,Hofmann:2020,Fujii:2020}.

We can calculate $\zeta_{\fermion}$ using the resummation method in
Ref.~\cite{Fujii:2021}, so that $\zeta_{\fermion}$ is provided by
\begin{equation}
\zeta_{\fermion}
=2\beta\int_{\bp}e^{-\beta\epsilon_{\bp}}
 \tilde{\Pi}^{F}_{RA}(\bp)\varphi(\bp)
+O(z^{3}),
\xlabel{eq:result-bulk-pinch}
\end{equation}
where we introduced the rescaled on-shell vertex function
\begin{equation}
\varphi(\bp)\equiv \frac{z\Gamma^{F}_{RA}(\bp)}{-2\Im[\Sigma^{R}(\bp)]},
\xlabel{eq:fermi-deviation-function}
\end{equation}
and employed shorthand notations for on-shell $\tilde{\Pi}^{F}_{RA}(\bp)\equiv \tilde{\Pi}^{F}(\epsilon_{\bp}-\mu+i0^{+},\epsilon_{\bp}-\mu-i0^{+};\bp), \Gamma^{F}_{RA}(\bp)\equiv \Gamma^{F}(\epsilon_{\bp}-\mu+i0^{+},\epsilon_{\bp}-\mu-i0^{+};\bp)$, and $\Sigma^{R}(\bp)\equiv\Sigma(\epsilon_{\bp}-\mu+i0^{+},\bp)$.
Since $\varphi(\bp)$ is $O(z)$ we find that $\zeta_{\fermion}$ is of the same order as $\zeta_{\pair}$.The resummation for the fermion pinch singularity is achieved by solving the self-consistent equation for the vertex function depicted in the upper-right panel of  Fig.~\ref{fig:fermion-bubble}.
The self-consistent equation for the vertex function is reduced into the linearized Boltzmann equation for $\varphi(\bp)$ as
\begin{equation}
\tilde{\Pi}^{F}_{RA}(\bp)
= \int_{\bq,\bp^{\prime},\bq^{\prime}}
 e^{-\beta\epsilon_{\bq}}
 W(\bp,\bq\vert \bp^{\prime},\bq^{\prime})
 \Bigl[
  \varphi(\bp) + \varphi(\bq)
  - \varphi(\bp^{\prime}) - \varphi(\bq^{\prime})
 \Bigr],
\xlabel{eq:linearized-Boltzmann}
\end{equation}
Here, $W(\bp,\bq\vert \bp^{\prime},\bq^{\prime})$ is the transition rate from initial $(\bp^{\prime},\bq^{\prime})$ to final momenta $(\bp,\bq)$ provided by
\begin{equation}
\begin{split}
W(\bp,\bq\vert \bp^{\prime},\bq^{\prime})
& = \vert D(\epsilon_{\bp}+\epsilon_{\bq}-2\mu+i0^{+},\bp+\bq)\vert ^2
 (2\pi)^{d+1}\\
 &\quad\times\delta(\epsilon_{\bp}+\epsilon_{\bq}-\epsilon_{\bp^{\prime}}-\epsilon_{\bq^{\prime}})
\delta(\bp+\bq-\bp^{\prime}-\bq^{\prime}).
\end{split}\xlabel{eq:transition}
\end{equation}

In order for the linearized Boltzmann Eq.~\eqref{eq:linearized-Boltzmann} to have a solution, the on-shell vertex function $\tilde{\Pi}^{F}_{RA}(\bp)$ must be orthogonal to $1$ and $\epsilon_{\bp}$, which correspond to the conserved particle number and the energy, respectively.
Here, the orthogonality is defined in terms of the inner product of $\langle A,B\rangle\equiv\int_{\bp}e^{-\beta\epsilon_{\bp}}A(\bp)B(\bp)$.
From the viewpoint of the kinetic theory, the orthogonality to each is necessary for $\varphi(\bp)$ to represent fluctuations around an equilibrium state with fixed particle number and energy.
We can confirm that the orthogonality relations are satisfied as
\begin{subequations}
\begin{align}
\int_{\bp}e^{-\beta\epsilon_{\bp}}\tilde{\Pi}^{F}_{RA}(\bp)
& = \int_{\bp}e^{-\beta\epsilon_{\bp}}\frac{C^{F}_{RA}(\bp)}{d\Omega_{d-1}ma^{d-2}}
-\frac{4z}{d\lambda_{T}^{d}}\pd{}{\beta}\delta b^{(\textrm{scat})}_{2}(\beta,a)
= 0, \\
\int_{\bp}e^{-\beta\epsilon_{\bp}}\epsilon_{\bp} \tilde{\Pi}^{F}_{RA}(\bp)
& = \int_{\bp}e^{-\beta\epsilon_{\bp}}\epsilon_{\bp}
    \frac{C^{F}_{RA}(\bp)}{d\Omega_{d-1}ma^{d-2}}
\nonumber\\
&\quad-\frac{z}{\beta\lambda_{T}^{d}}
\biggl[
 \pd{}{\beta}\delta b^{(\textrm{scat})}_{2}(\beta,a)
 - \frac{2}{d}\beta\pd{\mbox{}^{2}}{\beta^{2}}\delta b^{(\textrm{scat})}_{2}(\beta,a)
\biggr]
= 0,
\end{align}
\xlabel{eq:Chapman-Enskog-condition}
\end{subequations}
with $C^{F}_{RA}(\bp)\equiv C^{F}(\epsilon_{\bp}-\mu+i0^{+},\epsilon_{\bp}-\mu-i0^{+};\bp)$.
Here, in each first equality, we substituted 
Eq.~\eqref{eq:pressure-fluct}
into the pressure fluctuations of Eq.~\eqref{eq:fermion-vertex} and used the fact that the $\beta$ dependence of $b_2(\beta,a)$ in Eq.~\eqref{eq:b2virial} is contained in $\delta b^{(\textrm{scat})}_{2}(\beta,a)$ at a negative scattering length.
Also, in each second equality, the first term after integrating over all except the radial direction of the relative momentum is equal to the derivatives of $\delta b^{(\textrm{scat})}_{2}(\beta,a)$.
Therefore, the pressure fluctuations, i.e.,~the second and third terms in Eq.~\eqref{eq:fermion-vertex}, are essential for ensuring orthogonality and making the integrals~\eqref{eq:Chapman-Enskog-condition} vanish.
Note that these integrals become nonzero at positive scattering lengths because, in addition to the appearance of the pair pinch singularity, the second virial coefficient in the pressure fluctuation terms in Eq.~\eqref{eq:fermion-vertex} has the bound-state contribution.
This means that the resummation discussed in this section is valid only as far as the bound-state contributions are negligible.

In particular, $\tilde{\Pi}^{F}_{RA}(\bp)$ for $d=3$ near the unitary limit has the asymptotic form of
\begin{equation}
\tilde{\Pi}^{F}_{RA}(\bp)
= \frac{z}{\pi \beta}\frac{\lambda_{T}}{a}
\biggl[
 \frac{4}{3}\frac{F_{D}(\sqrt{\beta\epsilon_{\bp}})}{\sqrt{\beta\epsilon_{\bp}}}
 -1
 +\frac{2}{9}\beta\epsilon_{\bp}
\biggr]
+ O(z^{2},a^{-2}),
\xlabel{eq:DS-Xp}
\end{equation}
with $F_{D}(x)$ being the Dawson function and the use of the asymptotic form of the second virial coefficient $b_{2}=3/(4\sqrt{2})+\lambda_{T}/(\pi a)+O(a^{-2})$.
In two dimensions, although the scattering length is always positive, near the free-fermion limit ($\epsilon_B\to 0$), where the bound-state contribution is negligible, we can
 use the asymptotic expansion of the second virial
coefficient~\cite{Chafin:2013},
\begin{equation}
b_{2}=-\frac{1}{4}-\frac{1}{\ln(\beta\epsilon_{B})}+\frac{\gamma}{[\ln(\beta\epsilon_{B})]^{2}}+O([\ln(\beta\epsilon_{B})]^{-3}),
\xlabel{eq:approximated-b2-two-dim}
\end{equation}
with $\gamma=0.577\dots$ being Euler's constant, 
and
$\tilde{\Pi}^{F}_{RA}(\bp)$ is given by
\begin{equation}
\tilde{\Pi}^{F}_{RA}(\bp)
= z\frac{2}{\beta}\frac{1}{[\ln(\beta\epsilon_{B})]^{3}}
\Bigl[
 -2\textrm{Ei}(-\beta\epsilon_{\bp})+2\ln(\beta\epsilon_{\bp}/2)+1+2\gamma-\beta\epsilon_{\bp}
\Bigr]
+O(z^{2},[\ln(\beta\epsilon_{B})]^{-4}),
\xlabel{eq:CS-Xp}
\end{equation}
where $\textrm{Ei}(x)$ is the exponential integral.
Eq.~\eqref{eq:DS-Xp} is the same as $X_{p}$ in Ref.~\cite{Dusling:2013},
and Eq.~\eqref{eq:CS-Xp} is also the same as $X_{p}$ in
Ref.~\cite{Chafin:2013}.\footnote{Eq.~\eqref{eq:CS-Xp} is different in the sign from $X_{p}$ in
Ref.~\cite{Chafin:2013}, but it has no effect on the bulk viscosity.
Also, because the bound-state contribution $\delta b^{(\textrm{bound})}_{2}(\beta,a)$ is approximately independent of $\beta$ in Eq.~\eqref{eq:approximated-b2-two-dim},
$\tilde{\Pi}^{F}_{RA}(\bp)$ of Eq.~\eqref{eq:CS-Xp} has no bound-state contribution and results in the integrals in Eqs.~\eqref{eq:Chapman-Enskog-condition} being zero.}
Therefore, $\zeta_{\fermion}$ agrees with the kinetic result in
Refs.~\cite{Dusling:2013,Chafin:2013}.
Although $\zeta_{\pair}$ was calculated in
Refs.~\cite{Nishida:2019,Enss:2019,Hofmann:2020} and $\zeta_{\fermion}$ in
Refs.~\cite{Dusling:2013,Chafin:2013}, it is worth emphasizing that it is their
sum that provides the complete bulk viscosity when only the fermion pinch
singularity is relevant.

\subsection{The resulting bulk viscosity for $d=3$ at negative scattering length}

Let us compute the resulting bulk viscosity for $d=3$ at negative scattering length, where the pair pinch singularity is irrelevant.
First, substituting Eq.~\eqref{eq:scattering-continuum} into
Eq.~\eqref{eq:result-PB-bulk}, we obtain an analytic expression of $\zeta_{\pair}$
as~\cite{Enss:2019,Nishida:2019,Hofmann:2020}
\begin{equation}
\lambda_T^3\zeta_{\pair}
=z^2\frac{2\sqrt{2}}{9\pi}v^2
\Bigl[
    -1-(1+v^2)e^{v^2}\textrm{Ei}(-v^2)
\Bigr],
\xlabel{eq:result-bulk3Dpair-analitic}
\end{equation}
with $v\equiv \lambda_T/(\sqrt{2\pi}a)=\sqrt{\beta/m}a^{-1}$.
Next, to obtain $\zeta_{\fermion}$ we solve the linearized Boltzmann equation~\eqref{eq:linearized-Boltzmann} by expanding $\varphi(\bp)$ in terms of the generalized Laguerre polynomials $L^\alpha_n(x)$,
\begin{equation}
\varphi(\bp)=z\frac{\lambda_{T}}{a}\sum_{n=2}^{N+1}c_{n}L^{1/2}_{n}(\beta\epsilon_{\bp}).
\end{equation}
Here, the $n=0$ and $n=1$ terms are in the kernel of the collision operator and
vanish when substituted into Eqs.~\eqref{eq:result-bulk-pinch} and \eqref{eq:linearized-Boltzmann}.
Since even the simplest truncation $N=1$ provides a good approximation
as discussed in Refs.~\cite{Dusling:2013,Chafin:2013}, we employ the same
truncation and find
\begin{equation}
\varphi(\bp)=z\frac{\lambda_{T}}{a}\frac{c_{2}}{8}\Bigl(15+20\beta\epsilon_{\bp}+4(\beta\epsilon_{\bp})^{2}\Bigr)
\xlabel{eq:bulk3Dnegative-RTA}
\end{equation}
with
\begin{equation}
c_{2}
= \frac{1}{32\sqrt{2\pi}}
 \biggl(
  \int^{\infty}_{0}dx\,
   \frac{e^{-x}x^{3}}{\frac{1}{2\pi}(\frac{\lambda_{T}}{a})^{2}+x}
 \biggr)^{-1}
 \int^{\infty}_{0}dx\,
  \frac{e^{-x}\sqrt{x}}{\frac{1}{2\pi}(\frac{\lambda_{T}}{a})^{2}+x}
  (15-20x+4x^{2}).
\end{equation}
Substituting Eq.~\eqref{eq:bulk3Dnegative-RTA} into Eq.~\eqref{eq:result-bulk-pinch} leads to
\begin{equation}
\begin{split}
\lambda_{T}^{3}\zeta_{\fermion}
& = z^{2}\left(\frac{\lambda_{T}}{a}\right)^{2}
\frac{c_{2}}{24\pi^{3/2}}\int^{\infty}_{0}dx\,
 \frac{e^{-x}\sqrt{x}}{\frac{1}{2\pi}(\frac{\lambda_{T}}{a})^{2}+x}
 (15-20x+4x^{2}) \\
& = z^2\frac{v^2}{384\sqrt{2}}
    \frac{\left[2 \left(4+9 v^2 + 2 v^4\right)
 +\sqrt{\pi} e^{v^2} v \left(15+ 20v^2+4v^4\right) \text{erfc}(-v)\right]^2}{2- v^2+v^4+v^6 e^{v^2}\textrm{Ei}(-v^2)}
\end{split}\xlabel{eq:result-Fermi-pinch-bulk-RTA}
\end{equation}
with $v\equiv \lambda_T/(\sqrt{2\pi}a)=\sqrt{\beta/m}a^{-1}$.
This expression \eqref{eq:result-Fermi-pinch-bulk-RTA} provides the
contribution to the bulk viscosity from the fermionic kinetic theory for an arbitrary
negative scattering length and is an extension of the results near the unitary
limit in Ref.~\cite{Dusling:2013}.
The resulting bulk viscosities of Eqs.~\eqref{eq:result-bulk3Dpair-analitic} and \eqref{eq:result-Fermi-pinch-bulk-RTA} are plotted in  Fig.~\ref{fig:bulk3D-negative} as a function of the inverse scattering length.
As seen in  Fig.~\ref{fig:bulk3D-negative}, $\zeta_{\fermion}$ is sufficiently smaller than $\zeta_{\pair}$.
While $a^{2}\zeta_{\pair}$ diverges in the unitary limit, $a^{2}\zeta_{\fermion}$ converges to a finite value.

\section{Resummation of the pair pinch singularity}

\xlabel{sec:Resum-pair-pinch}
We next compute the bulk viscosity taking into account not only the fermion pinch singularity but also the pair one.
To consider the pair pinch singularity, we discuss the system with a positive scattering length hereafter that admits a paired bound state.

\subsection{Fermion--pair four point function}

Before we consider the bulk viscosity, let us discuss the four-point functions.
We classify the four-point functions depending on the propagators connected to the left and right sides of them, and focus on the following three types: fermion--fermion, pair--pair, and fermion--pair four-point functions, which are depicted by gray rectangles in the left, middle, and right of  Fig.~\ref{fig:def-of-order-of-four-pt-func}, respectively.
Since the parallel propagators can lead to pinch singularities, this classification is the key to considering the two types of the pinch singularities.
We show the following two properties of the fermion--pair four-point function $K_{\fp}(\ast)$: (i) $K_{\fp}(\ast)$ acquires an additional order in the iterations of the four-point functions, and (ii) $K_{\fp}(\ast)$ must be at least $O(z)$ for the propagators on both sides to have the pinch singularities simultaneously.

\subsubsection{Order counting for the iterated four-point functions}

In preparation for discussing at zero frequency, where the pinch singularities appear, we introduce the order of the four-point functions with respect to the fugacity at finite frequencies based on the order of the closed diagrams containing them.
We define the order of the four-point functions as the order of the correlation function that its minimum order is subtracted from, as shown in Fig.~\ref{fig:def-of-order-of-four-pt-func}.
When we refer to a fermion--fermion four-point function having $O(z^{n})$, we mean that the correlation function with that four-point function and $O(z^{0})$ vertices has $O(z^{n+1})$.
Here, since the fermion-bubble shape diagram without four-point functions has a minimum order of $z$, the order of the fermion-fermion four-point function and its closed diagram differ by one.
This order counting means that one order resulting from the propagator running backward is counted as that of the four-point function.
We count $K_{\fp}(\ast)$ as including one order of fugacity from the propagator running backward. 
The orders of the other four-point functions are defined similarly according to  Fig.~\ref{fig:def-of-order-of-four-pt-func}.

\begin{figure}[t]
\begin{align*}
\scalebox{1.2}{
\begin{tikzpicture}[scale=1,transform shape,baseline=-0.6ex]
\begin{feynhand}
    \vertex [squaredot] (a0) at (-1.9,0) {};
	\vertex (a2) at (-1.4,0.6);
	\vertex (a3) at (-0.4,0.6);
	\vertex (b2) at (-1.4,-0.6);
	\vertex (b3) at (-0.4,-0.6);
	\vertex [squaredot] (b0) at (0.1,0) {};
	\propag [fermion,looseness=0.7,quarter right] (b0) to (a3);
	\propag [fermion,looseness=0.7,quarter right] (a2) to (a0);
	\propag [fermion,looseness=0.7,quarter right] (b3) to (b0);
	\propag [fermion,looseness=0.7,quarter right] (a0) to (b2);
	\vertex [draw,fill=lightgray,scale=4,yscale=1.4] (c2) at (-0.9,0) {};
	\vertex [scale=0.9] (c2) at (-0.9,0) {$O(z^{n})$};
\end{feynhand}
\end{tikzpicture}
$\sim O(z^{n+1})\ \ $
\begin{tikzpicture}[scale=1,transform shape,baseline=-0.6ex]
\begin{feynhand}
	\vertex (a2) at (-1.4,0.6);
	\vertex (a3) at (-0.4,0.6);
	\vertex (b2) at (-1.4,-0.6);
	\vertex (b3) at (-0.4,-0.6);
	\vertex [squaredot] (a0) at (-2.0,0) {};
	\vertex [squaredot] (b0) at (0.2,0) {};
	\propag [double,double distance=0.4ex,thick,with arrow=0.5,arrow size=0.8em,looseness=0.7,quarter right] (b0) to (a3);
	\propag [double,double distance=0.4ex,thick,with arrow=0.5,arrow size=0.8em,looseness=0.7,quarter right] (a2) to (a0);
	\propag [double,double distance=0.4ex,thick,with arrow=0.5,arrow size=0.8em,looseness=0.7,quarter right] (b3) to (b0);
	\propag [double,double distance=0.4ex,thick,with arrow=0.5,arrow size=0.8em,looseness=0.7,quarter right] (a0) to (b2);
	\vertex [draw,fill=lightgray,scale=4,yscale=1.4] (c2) at (-0.9,0) {};
	\vertex [scale=0.9] (c2) at (-0.9,0) {$O(z^{n})$};
\end{feynhand}
\end{tikzpicture}
$\sim O(z^{n+2})\ \ $
\begin{tikzpicture}[scale=1,transform shape,baseline=-0.6ex]
\begin{feynhand}
	\vertex (a2) at (-1.4,0.6);
	\vertex (a3) at (-0.4,0.6);
	\vertex (b2) at (-1.4,-0.6);
	\vertex (b3) at (-0.4,-0.6);
	\vertex [squaredot] (a0) at (-1.9,0) {};
	\vertex [squaredot] (b0) at (0.2,0) {};
	\propag [double,double distance=0.4ex,thick,with arrow=0.5,arrow size=0.8em,looseness=0.7,quarter right] (b0) to (a3);
	\propag [fermion,looseness=0.7,quarter right] (a2) to (a0);
	\propag [double,double distance=0.4ex,thick,with arrow=0.5,arrow size=0.8em,looseness=0.7,quarter right] (b3) to (b0);
	\propag [fermion,looseness=0.7,quarter right] (a0) to (b2);
	\vertex [draw,fill=lightgray,scale=4,yscale=1.4] (c2) at (-0.9,0) {};
	\vertex [scale=0.9] (c2) at (-0.9,0) {$O(z^{n})$};
\end{feynhand}
\end{tikzpicture}
$\sim O(z^{n+2})$
}
\end{align*}
\caption{The gray rectangles on the left, middle, and right represent the fermion--fermion, pair--pair, and fermion--pair four-point functions, respectively.
We define that each four-point function has order $O(z^n)$ at finite frequencies when the corresponding correlation function with $O(z^0)$ vertices  (black squares) has the order written on the right-hand side of each.\xlabel{fig:def-of-order-of-four-pt-func}}
\end{figure}
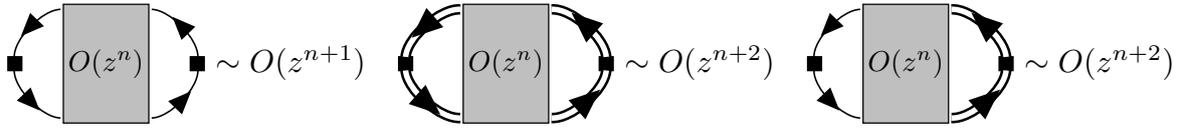

Based on the order counting method discussed in  Section~\ref{sec:order-counting-method} and the definition of the order of the four-point functions, we find that the iterations of the four-point functions contribute at a somewhat peculiar order shown in  Fig.~\ref{fig:iteration}.
Two iterations of the fermion--fermion or pair--pair four-point functions with $O(z^m)$ and $O(z^n)$ have the order of the product of theirs, i.e.,~$O(z^{m+n})$.
On the other hand, two iterations of the fermion--pair four-point function have the order $O(z^{m+n+1})$, which is one order greater than the order of their product.
This additional order in fugacity can be understood from the increase or decrease of the number of propagators running backward in imaginary time due to iterations of the four-point functions.

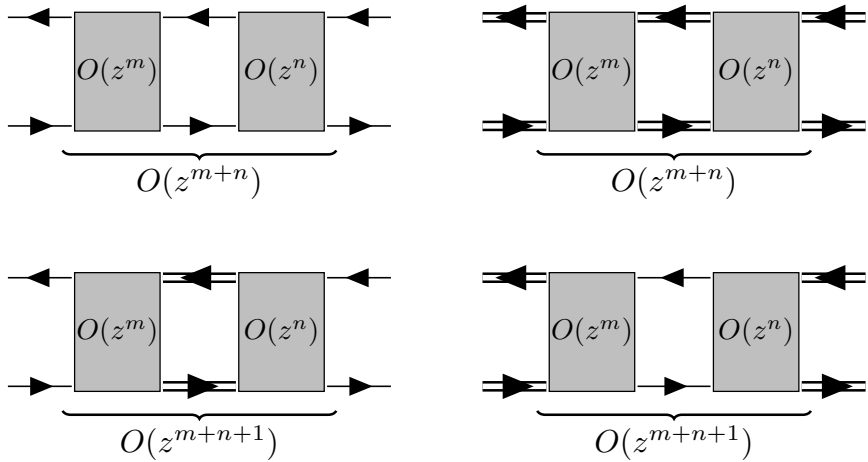
\begin{figure}[t]
\begin{align*}
\scalebox{1.2}{
\begin{tikzpicture}[scale=1,transform shape,baseline=-0.6ex]
\begin{feynhand}
	\vertex (a1) at (-2.1,0.6); \vertex (a2) at (-1.4,0.6);
	\vertex (a3) at (-0.4,0.6); \vertex (a4) at (0.4,0.6);
	\vertex (a5) at (1.4,0.6); \vertex (a6) at (2.1,0.6);
	\vertex (b1) at (-2.1,-0.6); \vertex (b2) at (-1.4,-0.6);
	\vertex (b3) at (-0.4,-0.6); \vertex (b4) at (0.4,-0.6);
	\vertex (b5) at (1.4,-0.6); \vertex (b6) at (2.1,-0.6);
	\propag [fermion] (a4) to (a3);
	\propag [fermion] (a6) to (a5);
	\propag [fermion] (a2) to (a1);
	\propag [fermion] (b3) to (b4);
	\propag [fermion] (b5) to (b6);
	\propag [fermion] (b1) to (b2);
	\vertex [draw,fill=lightgray,scale=4,yscale=1.4] (c1) at (0.9,0) {};
	\vertex [scale=0.9] (c1) at (0.9,0) {$O(z^{n})$};
	\vertex [draw,fill=lightgray,scale=4,yscale=1.4] (c2) at (-0.9,0) {};
	\vertex [scale=0.9] (c2) at (-0.9,0) {$O(z^{m})$};
	\vertex (d1) at (-1.5,-0.8); \vertex (d2) at (1.5,-0.8);
	\draw [thick,decoration={brace,mirror,raise=0.2em},decorate] (d1) -- (d2) node [pos=0.5,anchor=north,yshift=-0.3em] {$O(z^{m+n})$}; 
\end{feynhand}
\end{tikzpicture}
$\qquad$
\begin{tikzpicture}[scale=1,transform shape,baseline=-0.6ex]
\begin{feynhand}
	\vertex (a1) at (-2.1,0.6); \vertex (a2) at (-1.4,0.6);
	\vertex (a3) at (-0.4,0.6); \vertex (a4) at (0.4,0.6);
	\vertex (a5) at (1.4,0.6); \vertex (a6) at (2.1,0.6);
	\vertex (b1) at (-2.1,-0.6); \vertex (b2) at (-1.4,-0.6);
	\vertex (b3) at (-0.4,-0.6); \vertex (b4) at (0.4,-0.6);
	\vertex (b5) at (1.4,-0.6); \vertex (b6) at (2.1,-0.6);
	\propag [double,double distance=0.4ex,thick,with arrow=0.5,arrow size=0.8em] (a4) to (a3);
	\propag [double,double distance=0.4ex,thick,with arrow=0.5,arrow size=0.8em] (a6) to (a5);
	\propag [double,double distance=0.4ex,thick,with arrow=0.5,arrow size=0.8em] (a2) to (a1);
	\propag [double,double distance=0.4ex,thick,with arrow=0.5,arrow size=0.8em] (b3) to (b4);
	\propag [double,double distance=0.4ex,thick,with arrow=0.5,arrow size=0.8em] (b5) to (b6);
	\propag [double,double distance=0.4ex,thick,with arrow=0.5,arrow size=0.8em] (b1) to (b2);
	\vertex [draw,fill=lightgray,scale=4,yscale=1.4] (c1) at (0.9,0) {};
	\vertex [scale=0.9] (c1) at (0.9,0) {$O(z^{n})$};
	\vertex [draw,fill=lightgray,scale=4,yscale=1.4] (c2) at (-0.9,0) {};
	\vertex [scale=0.9] (c2) at (-0.9,0) {$O(z^{m})$};
	\vertex (d1) at (-1.5,-0.8); \vertex (d2) at (1.5,-0.8);
	\draw [thick,decoration={brace,mirror,raise=0.2em},decorate] (d1) -- (d2) node [pos=0.5,anchor=north,yshift=-0.3em] {$O(z^{m+n})$}; 
\end{feynhand}
\end{tikzpicture}
}
\end{align*}
\begin{align*}
\scalebox{1.2}{
\begin{tikzpicture}[scale=1,transform shape,baseline=-0.6ex]
\begin{feynhand}
	\vertex (a1) at (-2.1,0.6); \vertex (a2) at (-1.4,0.6);
	\vertex (a3) at (-0.4,0.6); \vertex (a4) at (0.4,0.6);
	\vertex (a5) at (1.4,0.6); \vertex (a6) at (2.1,0.6);
	\vertex (b1) at (-2.1,-0.6); \vertex (b2) at (-1.4,-0.6);
	\vertex (b3) at (-0.4,-0.6); \vertex (b4) at (0.4,-0.6);
	\vertex (b5) at (1.4,-0.6); \vertex (b6) at (2.1,-0.6);
	\propag [double,double distance=0.4ex,thick,with arrow=0.5,arrow size=0.8em] (a4) to (a3);
	\propag [fermion] (a6) to (a5);
	\propag [fermion] (a2) to (a1);
	\propag [double,double distance=0.4ex,thick,with arrow=0.5,arrow size=0.8em] (b3) to (b4);
	\propag [fermion] (b5) to (b6);
	\propag [fermion] (b1) to (b2);
	\vertex [draw,fill=lightgray,scale=4,yscale=1.4] (c1) at (0.9,0) {};
	\vertex [scale=0.9] (c1) at (0.9,0) {$O(z^{n})$};
	\vertex [draw,fill=lightgray,scale=4,yscale=1.4] (c2) at (-0.9,0) {};
	\vertex [scale=0.9] (c2) at (-0.9,0) {$O(z^{m})$};
	\vertex (d1) at (-1.5,-0.8); \vertex (d2) at (1.5,-0.8);
	\draw [thick,decoration={brace,mirror,raise=0.2em},decorate] (d1) -- (d2) node [pos=0.5,anchor=north,yshift=-0.3em] {$O(z^{m+n+1})$}; 
\end{feynhand}
\end{tikzpicture}
$\qquad$
\begin{tikzpicture}[scale=1,transform shape,baseline=-0.6ex]
\begin{feynhand}
	\vertex (a1) at (-2.1,0.6); \vertex (a2) at (-1.4,0.6);
	\vertex (a3) at (-0.4,0.6); \vertex (a4) at (0.4,0.6);
	\vertex (a5) at (1.4,0.6); \vertex (a6) at (2.1,0.6);
	\vertex (b1) at (-2.1,-0.6); \vertex (b2) at (-1.4,-0.6);
	\vertex (b3) at (-0.4,-0.6); \vertex (b4) at (0.4,-0.6);
	\vertex (b5) at (1.4,-0.6); \vertex (b6) at (2.1,-0.6);
	\propag [fermion] (a4) to (a3);
	\propag [double,double distance=0.4ex,thick,with arrow=0.5,arrow size=0.8em] (a6) to (a5);
	\propag [double,double distance=0.4ex,thick,with arrow=0.5,arrow size=0.8em] (a2) to (a1);
	\propag [fermion] (b3) to (b4);
	\propag [double,double distance=0.4ex,thick,with arrow=0.5,arrow size=0.8em] (b5) to (b6);
	\propag [double,double distance=0.4ex,thick,with arrow=0.5,arrow size=0.8em] (b1) to (b2);
	\vertex [draw,fill=lightgray,scale=4,yscale=1.4] (c1) at (0.9,0) {};
	\vertex [scale=0.9] (c1) at (0.9,0) {$O(z^{n})$};
	\vertex [draw,fill=lightgray,scale=4,yscale=1.4] (c2) at (-0.9,0) {};
	\vertex [scale=0.9] (c2) at (-0.9,0) {$O(z^{m})$};
	\vertex (d1) at (-1.5,-0.8); \vertex (d2) at (1.5,-0.8);
	\draw [thick,decoration={brace,mirror,raise=0.2em},decorate] (d1) -- (d2) node [pos=0.5,anchor=north,yshift=-0.3em] {$O(z^{m+n+1})$}; 
\end{feynhand}
\end{tikzpicture}
}
\end{align*}
\caption{Order counting of the iterated four-point functions at finite frequencies.
(Upper) Two iterations of the fermion--fermion or pair--pair four-point function have the combined order of the original four-point functions.
(Lower) Two iterations of the fermion--pair four-point function have an order that is greater by one than the combined order of the original four-point functions.\xlabel{fig:iteration}}
\end{figure}

\subsubsection{Simultaneous pinch singularities by propagators on both sides}

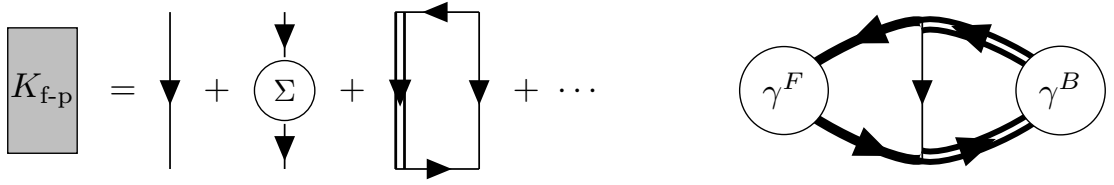
\begin{figure}[t]
\begin{tabular}{cc}
\begin{minipage}{0.64\hsize}
\begin{align*}
\scalebox{1.3}{
\begin{tikzpicture}[scale=1,baseline=-0.6ex]
\begin{feynhand}
	\vertex [draw,fill=lightgray,scale=3,yscale=1.82] (c1) at (0.46,0) {};
	\vertex (c1) at (0.46,0) {$K_{\fp}$};
\end{feynhand}
\end{tikzpicture}
$\ =\ $
\begin{tikzpicture}[scale=1,baseline=-0.6ex]
\begin{feynhand}
	\vertex (a1) at (0,-0.8); \vertex (a2) at (0,0.8);
	\propag [fermion] (a2) to (a1);
\end{feynhand}
\end{tikzpicture}
$\ +\ $
\begin{tikzpicture}[scale=1,baseline=-0.6ex]
\begin{feynhand}
	\vertex (a1) at (0,-0.8); \vertex (a2) at (0,-0.34);
	\vertex (a3) at (0,0.34); \vertex (a4) at (0,0.8);
	\propag [fermion] (a2) to (a1);
	\propag [fermion] (a4) to (a3);
	\vertex [draw,circle,scale=0.9] (a2) at (0,0) {$\Sigma$};
\end{feynhand}
\end{tikzpicture}
$\ +\ $
\begin{tikzpicture}[scale=1,baseline=-0.6ex]
\begin{feynhand}
	\vertex (a1) at (0,0.8); \vertex (a2) at (0,-0.8);
	\vertex (a3) at (0.8,0.8); \vertex (a4) at (0.8,-0.8);
	\vertex (b1) at (-0.05,0.8); \vertex (b2) at (-0.05,-0.8);
	\propag [double,double distance=0.4ex,thick,with arrow=0.5,arrow size=0.65em] (a1) to (a2);
	\propag [fermion] (a3) to (b1);
	\propag [fermion] (a3) to (a4);
	\propag [fermion] (b2) to (a4);
\end{feynhand}
\end{tikzpicture}
$\ +\ \cdots$
}
\end{align*}
\end{minipage}
\begin{minipage}{0.34\hsize}
\begin{align*}
\scalebox{1.3}{
\begin{tikzpicture}[baseline=-0.6ex]
\begin{feynhand}
	\vertex [circle] (b1) at (0,0); \vertex [circle] (b2) at (2.8,0);
	\vertex (d1) at (1.4,-0.7);
	\vertex (d2) at (1.4,0.7);
	\vertex (c1) at (1.38,-0.66);
	\vertex (c2) at (1.38,0.66);
	\propag [line width=1.0mm,quarter right,looseness=0.4,with arrow=0.34,arrow size=0.9em] (d2) to (b1);
	\propag [line width=1.0mm,quarter right,looseness=0.4,with arrow=0.66,arrow size=0.9em] (b1) to (d1);
	\propag [double,double distance=0.3ex,line width=0.4ex,with arrow=0.34,arrow size=0.9em,quarter right,looseness=0.5] (c1) to (b2);
	\propag [double,double distance=0.3ex,line width=0.4ex,with arrow=0.66,arrow size=0.9em,quarter right,looseness=0.5] (b2) to (c2);
	\propag [fermion] (d2) to (d1);
	\vertex [draw,fill=white,circle] (d1) at (0,0) {$\gamma^{F}$};
	\vertex [draw,fill=white,circle] (d2) at (2.8,0) {$\gamma^{B}$};
\end{feynhand}
\end{tikzpicture}
}
\end{align*}
\end{minipage}
\end{tabular}
\caption{(Left) Diagrammatic representation of the fermion--pair four-point function. The first term on the right-hand side is expressed by the thin line, i.e.,~the bare fermion propagator, and is the only $O(z^{0})$ contribution to the four-point function. The other diagrams are at least $O(z)$.
(Right) Diagrammatic representation of Eq.~\eqref{eq:def-chi-fp-0}.
While the vertical thin line represents the bare fermion propagator, the thick line and the thick double line represent the full fermion and pair propagators, respectively.
Also, $\gamma^{F}$ and $\gamma^{B}$ represent the vertex functions.\xlabel{fig:fermion-pair-four-pt-func}}
\end{figure}

The fermion--pair four-point function is composed of an infinite number of diagrams, as shown in the left panel of  Fig.~\ref{fig:fermion-pair-four-pt-func}.
Only the first term, i.e.,~the bare fermion propagator, is $O(z^0)$ as the fermion--pair four-point function, while the other terms are at least $O(z^1)$.
We show that the fermion--pair four-point function in $O(z^0)$ cannot have both the fermion and pair pinch singularities simultaneously on both sides of it.
In other words, when we take the correlation function depicted in the right panel of  Fig.~\ref{fig:fermion-pair-four-pt-func} as
\begin{equation}
\begin{split}
\chi_{\fp}(i\omega^{B})
&\equiv -\frac{1}{\beta^{2}}\sum_{m,n}\int_{\bp,\bq}
 \gamma^{F}(i\omega^{F}_{m}+i\omega^{B},i\omega^{F}_{m};\bp)
 \calG(i\omega^{F}_{m}+i\omega^{B},\bp)
 \calG(i\omega^{F}_{m},\bp) \\
&\quad \times
 G(i\omega^{B}_{n}-i\omega^{F}_{m},\bq-\bp)
 \calD(i\omega^{B}_{n}+i\omega^{B},\bq)
 \calD(i\omega^{B}_{n},\bq)
 \gamma^{B}(i\omega^{B}_{n}+i\omega^{B},i\omega^{B}_{n};\bq),
\end{split}\xlabel{eq:def-chi-fp-0}
\end{equation}
and the corresponding transport coefficient as
\begin{equation}
\sigma_{\fp}\equiv\lim_{\omega\to 0}\frac{\Im[\chi_{\fp}(\omega+i0^{+})]}{\omega},
\xlabel{eq:def-sigma0-fp}
\end{equation}
the transport coefficient $\sigma_{\fp}$ has no contribution in which the propagators, represented by the bold lines in the figure, lead to the two pinch singularities simultaneously.
(In~\xref{app-sec:sigmaFP-detailed-computation}, we calculate $\sigma_{\fp}$ specifically and show that $\sigma_{\fp}$ indeed has no contribution with the simultaneous fermion and pair pinch singularities.)
This is understood from the fact that the energy of a bound molecule enhanced by the pair pinch singularity differs from its on-shell energy by the binding energy $\epsilon_{B}$.
When the propagators on both sides of the fermion--pair four-point function have the pinch singularities simultaneously, this binding energy is propagated in the four-point function and is off-shell.
The spectrum of the bare fermion propagator taken as the four-point function in $\sigma_{\fp}$ does not have a width, so that off-shell energy cannot be propagated and thus, two pinch singularities do not occur simultaneously at $O(z^0)$. However,
if one considers other diagrams for the four-point function such as the ones depicted in Fig.~\ref{fig:fermion-pair-four-pt-func}, the transport coefficient can have contributions with two pinch singularities occurring simultaneously, albeit at higher order.

Therefore, an important lesson to be learned can be summarized as follows:
The fermion--pair four-point function at $O(z^0)$ does not allow the propagators on both sides of it to have the pinch singularities simultaneously.
In other words, for two types of the pinch singularities to occur simultaneously, there must be a four-point function of at least $O(z)$ inserted between them.

\begin{figure}[t]
\begin{tabular}{cc}
\begin{minipage}{0.3\hsize}
\begin{align*}
\scalebox{1.2}{
\begin{tikzpicture}[scale=1,transform shape,baseline=-0.6ex]
\begin{feynhand}
	\vertex [dot] (a1) at (0,0) {};
	\vertex [crossdot] (a2) at (2,0) {};
	\propag [double,double distance=0.3ex,line width=0.5ex,with arrow=0.55,arrow size=1.0em,half right,looseness=1.0] (a1) to (a2);
	\propag [double,double distance=0.3ex,line width=0.5ex,with arrow=0.45,arrow size=1.0em,half right,looseness=1.0] (a2) to (a1);
	\vertex [circle,fill=white] (a1) at (0,0) {};
	\vertex [crossdot] (a1) at (0,0) {};
	\vertex [draw,fill=lightgray,circle] (b2) at (2,0) {$\Gamma^{B}$};
\end{feynhand}
\end{tikzpicture}
}
\end{align*}
\end{minipage}
\begin{minipage}{0.66\hsize}
\begin{center}
\includegraphics[width=0.9\linewidth]{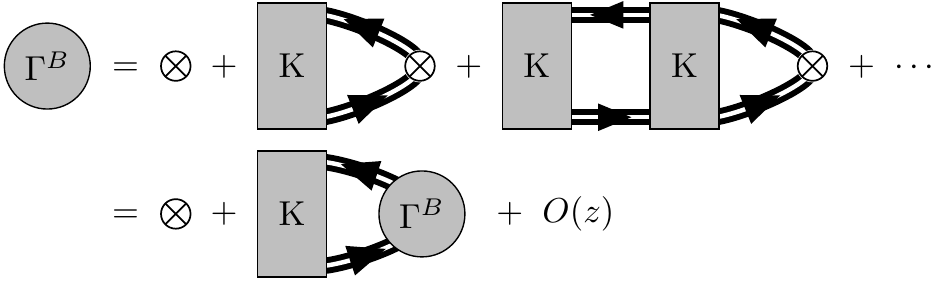}
\end{center}
\end{minipage}
\end{tabular}
\caption{(Left) Diagrammatic representation of Eq.~\eqref{eq:formal-contact-correl}, which is the imaginary-time-ordered correlation function for the contact. The gray bulb denotes the full vertex function.
(Right) Diagrammatic representation of the vertex function $\Gamma^{B}$ in terms of the pair--pair and fermion--pair four-point functions, $K_{\pp}$ and $K_{\fp}$. While the four-point functions do not contain propagators leading to the pinch singularity, the explicitly drawn propagators lead to the pinch singularity.\xlabel{fig:pair-self-consistent}}
\end{figure}

\subsection{Contact correlation function}

We consider the contact correlation function to find the pair pinch singularity.
Because the pair pinch singularity appears from the product of the pair propagators with the same frequency and momentum, the pair-bubble diagram, i.e.,~the first term in  Fig.~\ref{fig:contact-correlation}, has this singularity.
Looking ahead to the resummation, we express the contact correlation function formally as
\begin{equation}
\chi_{C}(i\omega^{B})
=\frac{1}{\beta}\sum_{m}\int_{\bp}
 m^{2}
 \calD(i\omega^{B}_{m}+i\omega^{B},\bp)
 \calD(i\omega^{B}_{m},\bp)
 \Gamma^{B}(i\omega^{B}_{m}+i\omega^{B},i\omega^{B}_{m};\bp),
\xlabel{eq:formal-contact-correl}
\end{equation}
which is represented diagrammatically in the left panel of  Fig.~\ref{fig:pair-self-consistent}.
Here, $\Gamma^{B}(i\omega^{B}_{m}+i\omega^{B},i\omega^{B}_{m};\bp)$ is the full vertex function to be determined in the following.

Replacing the Matsubara frequency summation with the complex contour
integration over $i\omega^{B}_{m}\to w$, one finds that the integrand can have the
singularities along $\Im[w]=0,-\omega^{B}$~\cite{Fujii:2021,Eliashberg:1962}.
Deforming the integral contour into four lines along $\Im[w]=\pm 0^{+},-\omega^{B}\pm 0^{+}$, we can evaluate the contour integration, and then the analytic continuation of $i\omega^{B}\to \omega+i0^{+}$ leads to
\begin{equation}
\begin{split}
\chi_{C}(\omega+i0^{+})
& = m^{2}\int_{\mathbb{R}\backslash\{0\}}\frac{d\varepsilon}{2\pi i}\int_{\bp}f_{B}(\varepsilon) \\
&\quad\times\Bigl[
 \calD^{R}(\varepsilon+\omega,\bp)
 \calD^{R}(\varepsilon,\bp)
 \Gamma^{B}(\varepsilon+\omega+i0^{+},\varepsilon+i0^{+};\bp) \\
&\qquad -
 \calD^{R}(\varepsilon+\omega,\bp)
 \calD^{A}(\varepsilon,\bp)
 \Gamma^{B}(\varepsilon+\omega+i0^{+},\varepsilon-i0^{+};\bp) \\
&\qquad +
 \calD^{R}(\varepsilon,\bp)
 \calD^{A}(\varepsilon-\omega,\bp)
 \Gamma^{B}(\varepsilon+i0^{+},\varepsilon-\omega-i0^{+};\bp)  \\
&\qquad -
 \calD^{A}(\varepsilon,\bp)
 \calD^{A}(\varepsilon-\omega,\bp)
 \Gamma^{B}(\varepsilon-i0^{+},\varepsilon-\omega-i0^{+};\bp)
\Bigr].
\end{split}\xlabel{eq:decomposed-contact-correlation}
\end{equation}
The second and third terms in Eq.~\eqref{eq:decomposed-contact-correlation} contain the pair pinch singularity in the zero-frequency limit, so that they have one order less in fugacity than the other terms and provide the dominant contribution.
Therefore, by applying the pair pinch singularity~\eqref{eq:pair-pinch} and keeping only the leading-order contribution, we obtain
\begin{equation}
\frac{1}{(d\Omega_{d-1}ma^{d-2})^{2}}
\lim_{\omega\to 0}
\frac{\Im[\chi_{C}(\omega+i0^{+})]}{\omega}
= \frac{\beta z^{2}e^{\beta\epsilon_{B}}}{d^{2}\Omega_{d-1}m^{2}a^{d}}
 \int_{\bp}
 e^{-\beta\epsilon_{\bp}/2}
 \frac{\Gamma^{B}_{RA\textrm{-pair}}(\bp)}{-\Im[\Delta^{R\textrm{-pair}}(\bp)]}+O(z^{2}),
\xlabel{eq:pair-pinch-bulk}
\end{equation}
where we introduced shorthand notations for pairing-shell $\Gamma^{B}_{RA\textrm{-pair}}(\bp)\equiv\gamma^{B}(\epsilon_{\bp}/2-2\mu-\epsilon_{B}+i0^{+},\epsilon_{\bp}/2-2\mu-\epsilon_{B}-i0^{+};\bp)$ and $\Delta^{R\textrm{-pair}}(\bp)\equiv\Delta(\epsilon_{\bp}/2-2\mu-\epsilon_{B}+i0^{+},\bp)$.
Eq.~\eqref{eq:pair-pinch-bulk} has an $O(z)$ contribution due to $\Gamma^{B}_{RA\textrm{-pair}}(\bp)\sim O(z^{0})$ and $\Im[\Delta^{R\textrm{-pair}}(\bp)]\sim O(z)$.
Here, the pressure fluctuation terms in Eq.~\eqref{eq:modified-stress} are $O(z)$ and therefore negligible compared to the vertex of $O(z^{0})$, so that Eq.~\eqref{eq:pair-pinch-bulk} provides the complete leading-order bulk viscosity.
Note that, as pointed out in
Refs.~\cite{Nishida:2019,Enss:2019,Hofmann:2020,Fujii:2020}, the spectral
function of the bulk viscosity has a term proportional to $\theta(a)\delta(\omega)$ at
$O(z^{2})$.
The delta function peak is caused by the pair pinch singularity and is removed by taking the broadening of the bound-state peak into account, leading to the $O(z)$ contribution discussed here.

\subsection{Self-consistent equation}

\xlabel{sec:self-consistent-eq}
Our remaining task is determining the full vertex function to the lowest order in fugacity.
If the naive expansion with respect to the fugacity is applied at nonzero frequencies, its
leading term is simply the bare vertex function, which is of $O(z^{0})$.
The other vertex functions are at least $O(z)$ at nonzero frequencies, and those which become $O(z^0)$ by the pinch singularity in the zero-frequency limit must be resummed.
When we write those propagators that lead to the pinch singularity explicitly in the diagrams and include all other propagators within the four-point functions, $\Gamma^{B}$ is expanded as shown in the top row of the right panel of  Fig.~\ref{fig:pair-self-consistent}.
Here, the pair--pair and fermion--pair four-point functions, $K_{\pp}$ and $K_{\fp}$, do not include the pinch singularity.
For example, the box diagram in which the fermion propagators do not lead to the pinch singularity is included in $K_{\pp}$.
Note that the fermion--pair four-point function is at least $O(z)$ for the propagators on its left and right sides to have the pinch singularity.
Then, diagrams such as the third term on the right-hand side in the right panel of  Fig.~\ref{fig:pair-self-consistent} cannot become $O(z^{0})$, even when one considers the pinch singularity.
This is because the iteration of the fermion--pair four-point function $K_{\fp}$ has order greater than the product of the original four-point functions, as seen in  Fig.~\ref{fig:iteration}.
Therefore, we only need to consider the resummation of the pair pinch singularities with $K_{\textrm{p-p}}$, and such a resummation can formally be achieved by replacing the bare vertex function with the full one, as shown in the bottom row of the right panel of  Fig.~\ref{fig:pair-self-consistent}.

The bottom row of the right panel of  Fig.~\ref{fig:pair-self-consistent} is expressed as
\begin{align}
\Gamma^{B}(i\omega^{B}_{m}+i\omega^{B},i\omega^{B}_{m};\bp)
& = m^{2}
 + \frac{1}{\beta} \sum_{m^{\prime}} \int_{\bp^{\prime}}
 K_{\pp}(i\omega^{B}_{m}+i\omega^{B},i\omega^{B}_{m};\bk\vert i\omega^{B}_{m^{\prime}}+i\omega^{B},i\omega^{B}_{m^{\prime}};\bp^{\prime}) \nonumber\\
& \quad \times
 \calD(i\omega^{B}_{m^{\prime}}+i\omega^{B},\bp^{\prime})
 \calD(i\omega^{B}_{m^{\prime}},\bp^{\prime})
 \Gamma^{B}(i\omega^{B}_{m^{\prime}}+i\omega^{B},i\omega^{B}_{m^{\prime}};\bp^{\prime}),
\xlabel{eq:pair-self-consistent}
\end{align}
where $K_{\pp}(\ast)$ has to be $O(z)$ for the lowest-order vertex function.
This is the closed equation that self-consistently determines the vertex function to lowest order in fugacity.

\subsection{Towards a kinetic theory}

\xlabel{sec:kinetic-theory}
For further computations on the self-consistent equation~\eqref{eq:pair-self-consistent}, we need a specific expression for $K_{\pp}(\ast)$.
However, the four-point function $K_{\pp}(\ast)$ is provided from infinite series of diagrams even for $O(z)$. In order to derive a kinetic equation as in the derivation of the linearized Boltzmann equation, we employ an approximation that replaces $K_{\pp}(\ast)$ with
\begin{align}
&K_{\rmbox}(i\omega^{B}_{m}+i\omega^{B},i\omega^{B}_{m};\bp\vert i\omega^{B}_{m^{\prime}}+i\omega^{B},i\omega^{B}_{m^{\prime}};\bp^{\prime}) \nonumber \\
& = -\frac{2}{\beta}\sum_{n}\int_{\bq} 
 G(i\omega^{B}_{m}-i\omega^{F}_{n},\bp-\bq)
 G(i\omega^{F}_{n}+i\omega^{B},\bq)
 G(i\omega^{F}_{n},\bq)
 G(i\omega^{B}_{m^{\prime}}-i\omega^{F}_{n},\bp^{\prime}-\bq).
\xlabel{eq:box-vertex-func}
\end{align}
Here, $K_{\rmbox}(\ast)$ is the four-point function inserted between the pair propagators in the box diagram (right part of Fig.~\ref{fig:contact-correlation}).
We evaluate the self-consistent equation with the above approximation so that it can be analytically continued into the equation for $\Gamma^{B}_{RA\textrm{-pair}}(\bp)$ needed to compute the transport coefficients according to Eq.~\eqref{eq:pair-pinch-bulk}.
According to the above approximation, we write the self-consistent equation as
\begin{equation}
\begin{split}
\Gamma^{B}(i\omega^{B}_{m}+i\omega^{B},i\omega^{B}_{m};\bp)
&= m^2+\frac{1}{\beta} \sum_{m^{\prime}} \int_{\bp^{\prime}}
 K_{\rmbox}(i\omega^{B}_{m}+i\omega^{B},i\omega^{B}_{m};\bp\vert i\omega^{B}_{m^{\prime}}+i\omega^{B},i\omega^{B}_{m^{\prime}};\bp^{\prime}) \\
&\quad\times
 \calD(i\omega^{B}_{m^{\prime}}+i\omega^{B},\bp^{\prime})
 \calD(i\omega^{B}_{m^{\prime}},\bp^{\prime})
 \Gamma^{B}(i\omega^{B}_{m^{\prime}}+i\omega^{B},i\omega^{B}_{m^{\prime}};\bp^{\prime}).
\end{split}
\end{equation}
Its analytic continuation with $i\omega^{F}_{m}\to \epsilon_{\bp}/2-\epsilon_{\bp-\bq}-\epsilon_{B}-\mu-i0^{+}$ followed by $i\omega^{B}\to i0^{+}$ is found as (See~\xref{app-sec:calculation-of-Box-vertex} for the detailed derivation)
\begin{equation}
\Gamma^{B}_{RA\textrm{-pair}}(\bp)
= m^{2}
+ z\frac{\Omega_{d-1}}{m^{2}a^{4-d}}\int_{\bp^{\prime}}
 K_{\rmbox}(\bp\vert \bp^{\prime})
 \frac{\Gamma^{B}_{RA\textrm{-pair}}(\bp^{\prime})}{-\Im[\Delta^{R\textrm{-pair}}(\bp^{\prime})]},
 \xlabel{eq:self-consistent-pair-vertex}
\end{equation}
where we introduced the pairing-shell kernel $K_{\rmbox}(\bp\vert \bp^{\prime})$ for the box diagram as
\begin{equation}
K_{\rmbox}(\bp\vert \bp^{\prime})
\equiv
2\int_{\bk,\bk^{\prime}}
e^{-\beta\epsilon_{\bk}}
\frac{
        (2\pi)^{d+1}
     \delta(\epsilon_{\bp}/2+\epsilon_{\bk}-\epsilon_{\bp^{\prime}}/2-\epsilon_{\bk^{\prime}})
     \delta^{d}(\bp+\bk-\bp^{\prime}-\bk^{\prime})
    }
 {[2\epsilon_{\bp/2-\bk^{\prime}}+\epsilon_{B}]^{2}}.
\xlabel{eq:def-Kernel-box}
\end{equation}
In analogy with Eq.~\eqref{eq:fermi-deviation-function}, we introduce the rescaled pairing-shell pair vertex function
\begin{equation}
\varphi_{\textrm{pair}}(\bp)
\equiv \frac{\Omega_{d-1}}{m^{2}a^{4-d}}
\frac{z\Gamma^{B}_{RA\textrm{-pair}}(\bp)}{-\Im[\Delta^{R\textrm{-pair}}(\bp)]}
\xlabel{eq:pair-deviation-function}
\end{equation}
and express the self-consistent equation \eqref{eq:self-consistent-pair-vertex} as
\begin{equation}
m^{2}=
\frac{-\Im[\Delta^{R\textrm{-pair}}(\bp)]}{z}
\frac{m^{2}a^{4-d}}{\Omega_{d-1}}
\varphi_{\pair}(\bp)
- \int_{\bp^{\prime}}
 K_{\rmbox}(\bp\vert \bp^{\prime})\varphi_{\pair}(\bp^{\prime}).
\xlabel{eq:self-consistent-eq-for-phi}
\end{equation}
Here, we have employed the approximation for the pair self-energy in which $\calT_{3}(\ast)$ is truncated after the second term in the lower panel of  Fig.~\ref{fig:self-energy};
this approximation is consistent with replacing the four-point function $K_{\pp}(\ast)$ with $K_{\rmbox}(\ast)$.
This consistency can be understood from the fact that the box diagram of the second term in  Fig.~\ref{fig:contact-correlation} is identical to the self-energy diagram with the approximated pair self-energy $\tilde{\Delta}(i\omega^{B}_{m},\bp)$, shown in  Fig.~\ref{fig:approximated-self-energy}, except for the positions of the vertices.\footnote{
The pairing-shell pair self-energy from the first term of $\calT_{3}$ has no imaginary part, so that it does not contribute to the self-consistent equation~\eqref{eq:self-consistent-eq-for-phi}, see also~\xref{app-sec:pair-self-energy}.
For this reason,  Fig.~\ref{fig:approximated-self-energy} shows the approximated pair self-energy, in which only the second term in the lower panel of  Fig.~\ref{fig:self-energy} is taken as $\calT_{3}(\ast)$.
}
The imaginary part of the approximated pairing-shell pair self-energy is provided by
\begin{equation}
\begin{split}
&-\Im[\tilde{\Delta}^{R\textrm{-pair}}(\bp)]
 = z\frac{\Omega_{d-1}}{m^{2}a^{4-d}}
 \int_{\bp^{\prime}}K_{\rmbox}(\bp\vert \bp^{\prime})
 + z \frac{\Omega_{d-1}}{m^{2}a^{4-d}} V_{\rmbox}(\bp)+O(z^{2}),
\end{split}\xlabel{eq:truncated_self-energy}
\end{equation}
with
\begin{equation}
V_{\rmbox}(\bp)
\equiv 2\frac{m^{2}a^{4-d}}{\Omega_{d-1}}\int_{\bp^{\prime},\bk,\bk^{\prime}}
 e^{-\beta\epsilon_{\bk}}
 \frac{(2\pi)^{d}\delta^{d}(\bp+\bk-\bp^{\prime}-\bk^{\prime})}
  {[2\epsilon_{\bp/2-\bk^{\prime}}+\epsilon_{B}]^{2}}
 \theta(\varepsilon)\rho_{D}(\varepsilon)\vert _{\varepsilon=\epsilon_{\bp}/2+\epsilon_{\bk}-\epsilon_{\bp^{\prime}}/2-\epsilon_{\bk^{\prime}}-\epsilon_{B}},
\xlabel{eq:def-potential-box}
\end{equation}
where $\rho_{D}(\varepsilon)$ is the scattering continuum defined in Eq.~\eqref{eq:scattering-continuum} (See~\xref{app-sec:pair-self-energy} for the detailed derivation).
Here, $V_{\rmbox}(\bp)$ is the potential representing the scattering from incoming $(\epsilon_{\bp}/2-\epsilon_{B},\bp;\epsilon_{\bk},\bk)$ to outgoing $(\epsilon_{\bp^{\prime}}/2,\bp^{\prime};\epsilon_{\bk^{\prime}},\bk^{\prime})$ energies and momenta.
Therefore, we obtain the self-consistent equation for $\varphi_{\pair}(\bp)$ as
\begin{equation}
m^{2}
= \int_{\bp^{\prime}}
 K_{\rmbox}(\bp\vert \bp^{\prime})
 [\varphi_{\pair}(\bp)-\varphi_{\pair}(\bp^{\prime})]
+ V_{\rmbox}(\bp)\varphi_{\pair}(\bp)+O(z).
\xlabel{eq:pair-kinetic-eq}
\end{equation}
Once the solution of $\varphi_{\pair}(\bp)$ is determined, the bulk viscosity in Eq.~\eqref{eq:pair-pinch-bulk} is provided by
\begin{equation}
\zeta
=\frac{z\beta}{(d\Omega_{d-1}a^{d-2})^{2}}
 \int_{\bp}e^{-\beta(\epsilon_{\bp}/2-\epsilon_{B})}\varphi_{\pair}(\bp)+O(z^{2}).
\xlabel{eq:result-bulk-positive}
\end{equation}
In the analogy to the derivation of the linearized Boltzmann equation, the self-consistent equation~\eqref{eq:pair-kinetic-eq} can be considered to be the kinetic equation for the bound pairs.
In contrast to the Boltzmann equation describing the collision process from two to two fermions, our kinetic equation, in particular the first term on the right-hand side, describes the change of state from one to one bound pair. This process can be understood from the corresponding Feynman diagram (the right panel of  Fig.~\ref{fig:approximated-self-energy}) as breaking and recombining a bound pair.

\begin{figure}[t]
\begin{tabular}{cc}
\begin{minipage}{0.4\hsize}
\begin{align*}
\scalebox{1.3}{
\begin{tikzpicture}[scale=1,transform shape,baseline=-0.6ex]
\begin{feynhand}
	\vertex [crossdot] (b1) at (0,0) {};
	\vertex [crossdot] (b3) at (2.0,0) {};
	\propag [double,double distance=0.4ex,thick,with arrow=0.18,with arrow=0.82,arrow size=0.7em,half right,looseness=0.8] (b1) to (b3);
	\propag [double,double distance=0.4ex,thick,with arrow=0.5,arrow size=0.8em,half right, looseness=0.8] (b3) to (b1);
	\vertex [draw,circle,fill=white,scale=2.0] (b2) at (1.0,-0.6) {};
	\vertex [] (b2) at (1.0,-0.6) {$\tilde{\Delta}$};
\end{feynhand}
\end{tikzpicture}
}
\end{align*}
\end{minipage}
\begin{minipage}{0.56\hsize}
\begin{align*}
\scalebox{1.3}{
\begin{tikzpicture}[scale=1,transform shape,baseline=-0.6ex]
\begin{feynhand}
	\vertex (b1) at (0,0);
	\vertex [draw,circle] (b2) at (1.0,0) {$\tilde{\Delta}$};
	\vertex (b3) at (2.0,0);
	\propag [double,double distance=0.4ex,thick,with arrow=0.5,arrow size=0.8em] (b1) to (b2);
	\propag [double,double distance=0.4ex,thick,with arrow=0.5,arrow size=0.8em] (b2) to (b3);
\end{feynhand}
\end{tikzpicture}
$\ =\ $
\begin{tikzpicture}[scale=1,transform shape,baseline=2.8ex]
\begin{feynhand}
	\vertex (b1) at (0.5,0.045); \vertex (b2) at (1.2,0.045);
	\vertex (b3) at (2.4,0.045); \vertex (b4) at (3.1,0.045);
	\vertex (c1) at (1.2,0); \vertex (c2) at (2.4,0);
	\vertex (c3) at (1.2,0.9); \vertex (c4) at (2.4,0.9);
	\propag [fermion] (c1) to (c2);
	\propag [fermion] (c4) to (c2);
	\propag [double,double distance=0.4ex,thick,with arrow=0.5,arrow size=0.8em] (c3) to (c4);
	\propag [fermion] (c1) to (c3);
	\propag [double,double distance=0.4ex,thick,with arrow=0.5,arrow size=0.8em] (b1) to (b2);
	\propag [double,double distance=0.4ex,thick,with arrow=0.5,arrow size=0.8em] (b3) to (b4);
	\propag [fermion,half right,looseness=1.8] (c4) to (c3);
\end{feynhand}
\end{tikzpicture}
}
\end{align*}
\end{minipage}
\end{tabular}
\caption{(Left) Diagrammatic representation of the self-energy diagram of the contact correlation function.
(Right) Diagrammatic representation of the approximated pair self-energy $\tilde{\Delta}$.
The self-energy diagram with $\tilde{\Delta}$ is identical to the box diagram, depicted in the second term of  Fig.~\ref{fig:contact-correlation}, except for the positions of the vertices.\xlabel{fig:approximated-self-energy}}
\end{figure}
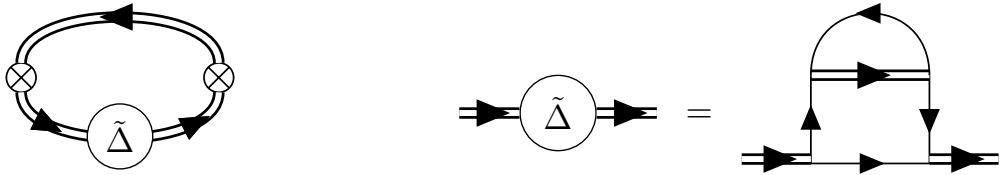

We comment on the bulk viscosity at positive scattering length.
By solving the obtained kinetic Eq.~\eqref{eq:pair-kinetic-eq} and computing Eq.~\eqref{eq:result-bulk-positive}, we can obtain the bulk viscosity at positive scattering length.
However, we do not present the value of the bulk viscosity because
Eq.~\eqref{eq:pair-kinetic-eq} is more complicated than the linearized Boltzmann equation and even a good approximation method such as the relaxation time approximation is unclear.
Note that although Eq.~\eqref{eq:result-bulk-positive} is proportional to $ze^{\beta\epsilon_B}a^{4-2d}$, it is controlled in the applicable regimes of the quantum virial expansion shown in  Fig.~\ref{fig:expansion-regime}.

\section{Summary}

\xlabel{sec:summary}
In this paper, we exactly evaluated the Kubo formula [Eq.~\eqref{eq:Kubo-formula}] for the bulk viscosity in the high-temperature limit to leading order in fugacity.
This task was achieved by summing up all contributions that are of higher order
in fugacity at nonzero frequencies but become comparable in the zero-frequency
limit due to the pinch singularity, as in the calculations for the shear
viscosity and the thermal conductivity~\cite{Fujii:2021}.
The key difference between the bulk viscosity and the other transport coefficients is that the bulk viscosity is expressed in terms of a correlation function of two-body operators, i.e.,~the contact operators [Eq.~\eqref{eq:def-contact-op}].
This difference not only makes the diagrams to be resummed more complicated than those for the other transport coefficients, but also brings about the emergence of the pair pinch singularity [Eq.~\eqref{eq:pair-pinch}] in addition to the fermion pinch singularity [Eq.~\eqref{eq:fermion-pinch}] even at the lowest order in fugacity.

For negative scattering lengths, where the pair pinch singularity does not appear, we showed that the bulk viscosity is reduced to the sum of two contributions: the pair-bubble contribution [Eq.~\eqref{eq:result-PB-bulk}] and the fermion kinetic contribution which is calculated from the linearized Boltzmann equation [Eqs.~\eqref{eq:result-bulk-pinch} and \eqref{eq:linearized-Boltzmann}].
Although both contribution were calculated separately in
Refs.~\cite{Dusling:2013,Chafin:2013,Nishida:2019,Enss:2019,Hofmann:2020,Fujii:2020},
we showed that only their sum provides the complete bulk viscosity.
We also obtained a new analytical expression for the bulk viscosity at arbitrary negative scattering length [Eqs.~\eqref{eq:result-bulk3Dpair-analitic} and \eqref{eq:result-Fermi-pinch-bulk-RTA}] and plotted the whole bulk viscosity for negative scattering lengths in three dimensions in  Fig.~\ref{fig:bulk3D-negative}.

For positive scattering lengths, we showed that  leading-order terms arise from contributions with the pair pinch singularity and need to be resummed.
Then, under the approximation that the irreducible four-point vertex is given by the box diagram, we reduced the self-consistent equation for the vertex function to a linearized equation for the pair distribution $\varphi_{\pair}(\bp)$ [Eq.~\eqref{eq:pair-kinetic-eq}].
The resulting linearized equation can be regarded as a kinetic equation for bound molecules in analogy to the linearized Boltzmann equation for fermions.
Although the numerical solution of this kinetic equation requires further work,
we identified the pair pinch singularity, which is responsible for the peaks
pointed out in previous
studies~\cite{Nishida:2019,Enss:2019,Hofmann:2020,Fujii:2020}, and provided a
resummation method for it.
We expect that this method will help to investigate the bulk viscosity in strongly correlated quantum many-body systems also beyond the high-temperature limit.

Finally, we comment on the divergence of the bulk viscosity multiplied by the squared scattering length, $a^{2}\zeta$, in the unitary limit, as seen in the right panel of  Fig.~\ref{fig:bulk3D-negative}.
This divergence originates from a singularity of the pair propagator in the unitary limit, which is different from the pair pinch singularity.
We briefly provide an intuitive understanding of this divergence.
The divergence of $a^{2}\zeta$ in the unitary limit arises from the integral $\int^{\infty}_{0}d\varepsilon\,[-\dif{f_B(\varepsilon)}{\varepsilon}]D^{R}(\varepsilon,\bp)D^{A}(\varepsilon,\bp)$, which appears in the calculation of the contact correlation function divided by $\omega$ in the zero-frequency limit.
Because of $D^{R}(\varepsilon,\bp)D^{A}(\varepsilon,\bp)=(4\pi/m)^{2}(m| \tilde{\varepsilon}| )^{-1}$ with $\tilde{\varepsilon}=\varepsilon-\epsilon_{\bp}/2+2\mu$ being the deviation from the on-shell energy, the integral diverges.
However, the broadening of the fermion spectrum at $O(z)$ leads in turn to a broadening $\chi\sim O(z)$ of the pair propagator of the same order via the diagram in  Fig.~\ref{fig:pair-propagator}. Hence, the product of the retarded and advanced pair propagators is replaced as
\begin{equation}
\int d\varepsilon\,
    \biggl[-\dif{f_B(\varepsilon)}{\varepsilon}\biggr]
    \calD^{R}(\varepsilon,\bp)
    \calD^{A}(\varepsilon,\bp)
\sim z^{2}\int_0^\infty \frac{d\tilde{\varepsilon}}{\sqrt{\tilde{\varepsilon}^2+\chi^2}}
\sim z^{2}\ln(1/z)
\end{equation}
and the integral is regularized for any value of $z$.
The logarithmic divergence of $z^{2}\ln(1/z)$ for $z\to0$ exactly at unitarity is consistent with the divergence of $a^{2}\zeta$ as unitarity is approached in the right panel of  Fig.~\ref{fig:bulk3D-negative}.
This singularity is interesting as a strong coupling effect because it is peculiar to the unitary limit, and we hope that a rigorous evaluation method for this singularity will be established in the future, as is the case for the pinch singularity.


\section*{Acknowledgments}
 The authors thank Jan Martin Pawlowski, Jeff Maki, and Masaru Hongo for valuable discussions.
This work is supported by the Deutsche Forschungsgemeinschaft (DFG, German Research Foundation), project-ID 273811115 (SFB1225 ISOQUANT) and under Germany's Excellence Strategy EXC2181/1-390900948 (the Heidelberg STRUCTURES Excellence Cluster).

\appendix
\section{Pair self-energy}
\xlabel{app-sec:pair-self-energy}
\subsection{Derivation of Eq.~\eqref{eq:pair-self-energy}}
The pair self-energy is provided by
\begin{align}
\Delta(i\omega^{B}_{m},\bp)
& = \frac{2}{\beta}\sum_{n}\int_{\bq}
 \calT_{3}(i\omega^{B}_{m},\bp;i\omega^{F}_{n},\bq\vert i\omega^{B}_{m},\bp;i\omega^{F}_{n},\bq)G(i\omega^{F}_{n},\bq) \nonumber\\
& \quad  - \left.\frac{2}{\beta}\sum_{n}\int_{\bq}
 G(i\omega^{B}_{m}-i\omega^{F}_{n},\bp-\bq)
 G(i\omega^{F}_{n},\bq)
 \right|_{O(z^0)~\textrm{part}},
\end{align}
where the second term serves to remove the contribution already included in the definition of the bare pair propagator $D(i\omega^{B}_{m},\bp)$.
We divide the pair-fermion scattering $T$-matrix $\calT_{3}$ into the first term and the other terms in the lower panel of  Fig.~\ref{fig:self-energy} and express it as
\begin{align}
\calT_{3}(i\omega^{B}_{m},\bp;i\omega^{F}_{n},\bq\vert i\omega^{B}_{m^{\prime}},\bp^{\prime};i\omega^{F}_{n^{\prime}},\bq^{\prime}) 
& = G(i\omega^{B}_{m}-i\omega^{F}_{n^{\prime}},\bp-\bq^{\prime})
\delta_{m+n,m^{\prime}+n^{\prime}} \delta(\bp+\bq-\bp^{\prime}-\bq^{\prime}) \nonumber\\
& \quad
+ \calT^{\prime}_{3}(i\omega^{B}_{m},\bp;i\omega^{F}_{n},\bq\vert i\omega^{B}_{m^{\prime}},\bp^{\prime};i\omega^{F}_{n^{\prime}},\bq^{\prime}).
\end{align}
Accordingly, the pair self-energy is divided into
\begin{equation}
\Delta(i\omega^{B}_{m},\bp)=\Delta^{(1)}(i\omega^{B}_{m},\bp)+\Delta^{(2)}(i\omega^{B}_{m},\bp)
\end{equation}
with
\begin{align}
\Delta^{(1)}(i\omega^{B}_{m},\bp)
& = \left.\frac{2}{\beta}\sum_{n}\int_{\bq}
 G(i\omega^{B}_{m}-i\omega^{F}_{n},\bp-\bq)
 G(i\omega^{F}_{n},\bq)
 \right|_{\textrm{removed}~O(z^{0})~\textrm{part}}, \\
\Delta^{(2)}(i\omega^{B}_{m},\bp)
& =
 \frac{2}{\beta}\sum_{n}\int_{\bq}
 \calT^{\prime}_{3}(i\omega^{B}_{m},\bp;i\omega^{F}_{n},\bq\vert i\omega^{B}_{m},\bp;i\omega^{F}_{n},\bq)G(i\omega^{F}_{n},\bq).
\xlabel{eq:def-Delta2}
\end{align}
The first term $\Delta^{(1)}(i\omega^{B}_{m},\bp)$ is calculated as
\begin{equation}
\Delta^{(1)}(i\omega^{B}_{m},\bp)
 = 2z\int_{\bq}e^{-\beta\epsilon_{\bq}}
[
 G(i\omega^{B}_{m}-\epsilon_{\bq}+\mu,\bp+\bq)
 + G(i\omega^{B}_{m}-\epsilon_{\bq}+\mu,\bp-\bq)
]+O(z^{2}).
\xlabel{eq:Delta1-calculated}
\end{equation}
In $\Delta^{(2)}(i\omega^{B},\bp)$, on the other hand, only the pole of the fermion propagator in Eq.~\eqref{eq:def-Delta2} leads to an $O(z)$ contribution, which is calculated as
\begin{equation}
\Delta^{(2)}(i\omega^{B}_{m},\bp)
=2z\int_{\bq}e^{-\beta\epsilon_{\bq}}
 \calT^{\prime}_{3}(i\omega^{B}_{m},\bp;\epsilon_{\bq}-\mu,\bq\vert i\omega^{B}_{m},\bp;\epsilon_{\bq}-\mu,\bq)+O(z^{2}).
 \xlabel{eq:Delta2-calculated}
\end{equation}
The sum of Eqs.~\eqref{eq:Delta1-calculated} and \eqref{eq:Delta2-calculated} provides Eq.~\eqref{eq:pair-self-energy}.

\subsection{Derivation of Eq.~\eqref{eq:truncated_self-energy}
}
We employ an approximation of truncating $\calT_{3}(\ast)$ up to the second term in  Fig.~\ref{fig:self-energy}.
Then, the approximated pair self-energy $\tilde{\Delta}(i\omega^{B}_{m},\bp)$ is divided into
\begin{equation}
\tilde{\Delta}(i\omega^{B}_{m},\bp)
=\Delta^{(1)}(i\omega^{B}_{m},\bp)+\tilde{\Delta}^{(2)}(i\omega^{B}_{m},\bp),
\end{equation}
where $\Delta^{(1)}(i\omega^{B}_{m},\bp)$ is given by Eq.~\eqref{eq:Delta1-calculated} and $\Delta^{(2)}(i\omega^{B}_{m},\bp)$ is approximated to $\tilde{\Delta}^{(2)}(i\omega^{B}_{m},\bp)$ by the truncation.
The second term of $\calT_{3}(\ast)$ in  Fig.~\ref{fig:self-energy} reads
\begin{align}
&{\calT}_{3}(i\omega^{B}_{m},\bp;i\omega^{F}_{n},\bq\vert i\omega^{B}_{m},\bp;i\omega^{F}_{n},\bq)\vert_{\text{second term}} \nonumber \\
&= - \frac{1}{\beta}\sum_{l}\int_{\bk}
 G(i\omega^{B}_{m}-i\omega^{F}_{l},\bp-\bk)
 G(i\omega^{F}_{l},\bk)^{2}
 D(i\omega^{F}_{n}+i\omega^{F}_{l},\bq+\bk),
\end{align}
and is calculated as
\begin{align}
&\calT_{3}(i\omega^{B}_{m},\bp;i\omega^{F}_{n},\bq\vert i\omega^{B}_{m},\bp;i\omega^{F}_{n},\bq)\vert_{\textrm{second term}} \nonumber \\
&= \int_{\bk}
 G(i\omega^{B}_{m}-\epsilon_{\bp-\bk}+\mu,\bk)^{2}
 D(i\omega^{B}_{m}+i\omega^{F}_{n}-\epsilon_{\bp-\bk}+\mu,\bq+\bk)+O(z).
\end{align}
Substituting this into Eq.~\eqref{eq:Delta2-calculated} leads to
\begin{equation}
\tilde{\Delta}^{(2)}(i\omega^{B}_{m},\bp)
= 2z \int_{\bq,\bk}e^{-\beta\epsilon_{\bq}}
 \frac{D(i\omega^{B}_{m}+\epsilon_{\bq}-\epsilon_{\bp-\bk},\bq+\bk)}
  {[i\omega^{B}_{m}-\epsilon_{\bp-\bk}-\epsilon_{\bk}+2\mu]^{2}}+O(z^{2}).
\end{equation}
The approximated pairing-shell pair self-energy is defined as $\tilde{\Delta}^{R\textrm{-pair}}(\bp)\equiv\tilde{\Delta}(\epsilon_{\bp}/2-2\mu-\epsilon_{B}+i0^{+},\bp)$, and its imaginary part is obtained as
\begin{equation}
-\Im[\tilde{\Delta}^{R\textrm{-pair}}(\bp)]
= 2z \int_{\bq,\bk}e^{-\beta\epsilon_{\bq}}
 \frac{-\Im[D(\epsilon_{\bp}/2+\epsilon_{\bq}-\epsilon_{\bp-\bk}-\epsilon_{B}-2\mu+i0^{+},\bq+\bk)]}
  {[2\epsilon_{\bk-\bp/2}+\epsilon_{B}]^{2}},
\end{equation}
where the imaginary part of $\Delta^{(1);R\textrm{-pair}}(\bp)$ vanishes to $O(z)$ because of
\begin{equation}
\Im[G(\epsilon_{\bp}/2-\epsilon_{\bq}-\epsilon_{B}-\mu+i0^{+},\bp\pm\bq)]=0.
\end{equation}
Finally, by using Eq.~\eqref{eq:pair-spectrum}, we arrive at Eq.~\eqref{eq:truncated_self-energy}.

\section{Transport coefficient with the $O(z^0)$ fermion--pair four-point function}
\xlabel{app-sec:sigmaFP-detailed-computation}

\subsection{Calculation of the transport coefficient}

In this appendix, we provide detailed calculations of $\sigma_{\fp}$ in Eq.~\eqref{eq:def-sigma0-fp}.
To calculate $\sigma_{\fp}$, let us start with $\chi_{\fp}(i\omega^{B})$ in Eq.~\eqref{eq:def-chi-fp-0}.
With the Matsubara frequency summation over $\omega^{F}_{m}$ replaced by the complex contour integration over $i\omega^{F}_{m}\to w$, the integrand of $\chi_{\fp}(i\omega^{B})$ has singularities only along $\Im[w] = 0, -\omega^{B}$ in addition to pole at $w=i\omega^{B}_{n}-\epsilon_{\bq-\bp}+\mu$ in the complex plane of $w$.
Therefore, we can deform its contour into four horizontal lines along $\Im[w] = \pm0^{+}, -\omega^{B}\pm0^{+}$ and clockwise circles around the pole.
Accordingly, we divide $\chi_{\fp}(i\omega^{B})$ and $\sigma_{\fp}$ into two parts: the contribution from the pole and the contribution from integrals on the four lines,
\begin{equation}
\chi_{\fp}(i\omega^{B})
=\chi_{\fp;\pole}(i\omega^{B})
+ \chi_{\fp;\lines}(i\omega^{B}),
\end{equation}
and $\sigma_{\fp}=\sigma_{\fp;\pole}+\sigma_{\fp;\lines}$ with
\begin{equation}
\sigma_{\fp;\pole}
=\lim_{\omega\to 0}\frac{\Im[\chi_{\fp;\pole}(\omega+i0^{+})]}{\omega},\qquad
\sigma_{\fp;\lines}
=\lim_{\omega\to 0}\frac{\Im[\chi_{\fp;\lines}(\omega+i0^{+})]}{\omega}.
\end{equation}
Evaluating the contour integrations, we find
\begin{equation}
\begin{split}
&\chi_{\fp;\pole}(i\omega^{B})
= \frac{1}{\beta}\sum_{n}\int_{\bp,\bq}f_{F}(-\epsilon_{\bq-\bp}+\mu) \\
&\times
 \gamma^{F}(i\omega^{B}_{n}+i\omega^{B}-\epsilon_{\bq-\bp}+\mu,i\omega^{B}_{n}-\epsilon_{\bq-\bp}+\mu;\bp)
 \calG(i\omega^{B}_{n}-\epsilon_{\bq-\bp}+\mu,\bp) \\
&\times
 \calG(i\omega^{B}_{n}+i\omega^{B}-\epsilon_{\bq-\bp}+\mu,\bp)
 \calD(i\omega^{B}_{n}+i\omega^{B},\bq)
 \calD(i\omega^{B}_{n},\bq)
 \gamma^{B}(i\omega^{B}_{n}+i\omega^{B},i\omega^{B}_{n};\bq),
\end{split}
\end{equation}
and
\begin{equation}
\begin{split}
&\chi_{\fp;\lines}(i\omega^{B}) \\
&= \frac{1}{\beta}\sum_{n}
 \int_{\bp,\bq}
 \int^{\infty}_{-\infty}\frac{d\varepsilon}{2\pi i}
 f_{F}(\varepsilon)
 \calD(i\omega^{B}_{n}+i\omega^{B},\bq)
 \calD(i\omega^{B}_{n},\bq)
 \gamma^{B}(i\omega^{B}_{n}+i\omega^{B},i\omega^{B}_{n};\bq) \\
&\quad\times
\Bigl[
 \gamma^{F}(\varepsilon+i\omega^{B},\varepsilon+i0^{+};\bp)
 \calG(\varepsilon+i\omega^{B},\bp)
 \calG(\varepsilon+i0^{+},\bp)
 G(i\omega^{B}_{n}-\varepsilon,\bq-\bp) \\
&\qquad
 -
 \gamma^{F}(\varepsilon+i\omega^{B},\varepsilon-i0^{+};\bp)
 \calG(\varepsilon+i\omega^{B},\bp)
 \calG(\varepsilon-i0^{+},\bp)
 G(i\omega^{B}_{n}-\varepsilon,\bq-\bp) \\
&\qquad
 +
 \gamma^{F}(\varepsilon+i0^{+},\varepsilon-i\omega^{B};\bp)
 \calG(\varepsilon+i0^{+},\bp)
 \calG(\varepsilon-i\omega^{B},\bp)
 G(i\omega^{B}_{n}+i\omega^{B}-\varepsilon,\bq-\bp) \\
&\qquad
 -
 \gamma^{F}(\varepsilon-i0^{+},\varepsilon-i\omega^{B};\bp)
 \calG(\varepsilon-i0^{+},\bp)
 \calG(\varepsilon-i\omega^{B},\bp)
 G(i\omega^{B}_{n}+i\omega^{B}-\varepsilon,\bq-\bp)
 \Bigr].
\end{split}
\end{equation}
Furthermore, after the replacement the Matsubara frequency summation over
$i\omega^{B}_{n}$ with the contour integration over $i\omega^{B}_{n}\to w$, the integrands
have the branches along $\Im[w]=0,-\omega^{B}$ in both $\chi_{\fp;\pole}(i\omega^{B})$ and
$\chi_{\fp;\lines}(i\omega^{B})$~\cite{Fujii:2021,Eliashberg:1962}.
Thus, one can evaluate the integration with the deformation of its contour into four straight lines along the branches as before.

Let us first consider $\sigma_{\fp;\pole}$, which is found as
\begin{equation}
\begin{split}
&\sigma_{\fp;\pole}
= -\int_{\bp,\bq}f_{F}(-\epsilon_{\bq-\bp}+\mu)
 \int^{\infty}_{-\infty} \frac{d\varepsilon}{2\pi}
 \dif{f_{B}(\varepsilon)}{\varepsilon} \\
&\times
 \gamma^{F}(\varepsilon-\epsilon_{\bq-\bp}+\mu+i0^{+},\varepsilon-\epsilon_{\bq-\bp}+\mu-i0^{+};\bp)
 \gamma^{B}(\varepsilon+i0^{+},\varepsilon-i0^{+};\bq) \\
&\times
 \calG^{R}(\varepsilon-\epsilon_{\bq-\bp}+\mu,\bp)
 \calG^{A}(\varepsilon-\epsilon_{\bq-\bp}+\mu,\bp)
 \calD^{R}(\varepsilon,\bq)
 \calD^{A}(\varepsilon,\bq)
 +O(z^{2}),
\end{split}\xlabel{eq:sigma0-fp-pole}
\end{equation}
where we supposed $\gamma^{F}(\ast)=\gamma^{B}(\ast)=O(z^{0})$ in the order counting.
In the omitted part as $O(z^{2})$ of $\sigma_{\fp;\pole}$, there is no product of the propagators which can lead to the pinch singularities, so that we do not discuss it.

Next, we consider $\sigma_{\fp;\lines}$, but it has no $O(z)$ contribution even when the pinch singularities are taken into account.
For this reason, we take only the contribution where both the product of the pair propagators and the product of the fermion ones can lead to the pinch singularities, as in Eq.~\eqref{eq:sigma0-fp-pole}.
Such contribution is straightforwardly found as
\begin{equation}
\begin{split}
&\sigma_{\fp;\lines}
\simeq \int_{\bp,\bq}
 \int^{\infty}_{-\infty}\frac{d\varepsilon}{2\pi}
 f_{F}(\varepsilon-\epsilon_{\bq-\bp}+\mu)
 \dif{f_{B}(\varepsilon)}{\varepsilon} \\
&\times
 \gamma^{F}(\varepsilon-\epsilon_{\bq-\bp}+\mu+i0^{+},\varepsilon-\epsilon_{\bq-\bp}+\mu-i0^{+};\bp)
 \gamma^{B}(\varepsilon+i0^{+},\varepsilon-i0^{+};\bq) \\
&\times
 \calG^{R}(\varepsilon-\epsilon_{\bq-\bp}+\mu,\bp)
 \calG^{A}(\varepsilon-\epsilon_{\bq-\bp}+\mu,\bp)
 \calD^{R}(\varepsilon,\bq)
 \calD^{A}(\varepsilon,\bq).
\end{split}\xlabel{eq:sigma0-fp-lines}
\end{equation}

The sum of Eqs.~\eqref{eq:sigma0-fp-pole} and \eqref{eq:sigma0-fp-lines} provides
\begin{equation}
\begin{split}
\sigma_{\fp}
&\simeq \beta\int_{\bp,\bq}
  \int^{\infty}_{-\infty} \frac{d\varepsilon}{2\pi}
  [f_{F}(-\epsilon_{\bq-\bp}+\mu)-f_{F}(\varepsilon-\epsilon_{\bq-\bp}+\mu)]
  f_{B}(\varepsilon)[1+f_{B}(\varepsilon)] \\
&\quad\times
 \calG^{R}(\varepsilon-\epsilon_{\bq-\bp}+\mu,\bp)
 \calG^{A}(\varepsilon-\epsilon_{\bq-\bp}+\mu,\bp)
 \calD^{R}(\varepsilon,\bq)
 \calD^{A}(\varepsilon,\bq) \\
&\quad\times
 \gamma^{F}(\varepsilon-\epsilon_{\bq-\bp}+\mu+i0^{+},\varepsilon-\epsilon_{\bq-\bp}+\mu-i0^{+};\bp)
 \gamma^{B}(\varepsilon+i0^{+},\varepsilon-i0^{+};\bq).
\end{split}\xlabel{eq:sigma0-fp}
\end{equation}

\subsection{Pinch singularities}

As we see in the following, Eq.~\eqref{eq:sigma0-fp} has no contribution with both the fermion and pair pinch singularities simultaneously, so that we consider the cases separately where there is the fermion or pair pinch singularity: $\sigma_{\fp}=\sigma_{\fp}\vert_{\textrm{fermi-pinch}}+\sigma_{\fp}\vert_{\textrm{pair-pinch}}$.

First, when the product of the fermion propagators leads to the pinch singularity \eqref{eq:fermion-pinch}, it turns into
\begin{equation}
\begin{split}
&\sigma_{\fp}\vert _{\textrm{fermi-pinch}}
= \beta\int_{\bp,\bq}
 e^{-\beta(\epsilon_{\bq-\bp}+\epsilon_{\bp})}
 \frac{z^{2}\gamma^{F}_{RA}(\bp)}{-2\Im[\Sigma^{R}(\bp)]}
 \calD^{R}(\epsilon_{\bq-\bp}+\epsilon_{\bp}-2\mu,\bq) \\
&\times
 \calD^{A}(\epsilon_{\bq-\bp}+\epsilon_{\bp}-2\mu,\bq)
 \gamma^{B}(\epsilon_{\bq-\bp}+\epsilon_{\bp}-2\mu+i0^{+},\epsilon_{\bq-\bp}+\epsilon_{\bp}-2\mu-i0^{+};\bq)
 +O(z^{2}),
\end{split}\xlabel{eq:sigma0-fermi-pinch}
\end{equation}
where we supposed $\gamma^{F}(\ast)=\gamma^{B}(\ast)=O(z^{0})$ in the order counting.
In this expression, the product of the pair propagators cannot have the pinch singularity because $\calD^{R}(\epsilon_{\bq-\bp}+\epsilon_{\bp}-2\mu,\bq)$ does not have the binding peak at $O(z^{0})$:
\begin{equation}
\Im[\calD^{R}(\epsilon_{\bq-\bp}+\epsilon_{\bp}-2\mu,\bq)]=-\rho_{D}(2\epsilon_{\bq/2-\bp})+O(z).
\end{equation}
The pair propagators in Eq.~\eqref{eq:sigma0-fermi-pinch} are simply replaced with the bare ones and thus we get, with $\gamma^{F}_{RA}(\bp)$ denoting the on-shell vertex defined in analogy with Eq.~\eqref{eq:fermi-deviation-function},
\begin{equation}
\begin{split}
&\sigma_{\fp}\vert _{\textrm{fermion-pinch}}
= \beta\int_{\bp,\bq}
 e^{-\beta(\epsilon_{\bq-\bp}+\epsilon_{\bp})}
 \frac{z^{2}\gamma^{F}_{RA}(\bp)}{-2\Im[\Sigma^{R}(\bp)]}
 \vert D(\epsilon_{\bq-\bp}+\epsilon_{\bp}-2\mu+i0^{+},\bq)\vert ^{2} \\
&\times
 \gamma^{B}(\epsilon_{\bq-\bp}+\epsilon_{\bp}-2\mu+i0^{+},\epsilon_{\bq-\bp}+\epsilon_{\bp}-2\mu-i0^{+};\bq)
 +O(z^{2}).
\end{split}\xlabel{eq:sigma0-fermi-pinch-result}
\end{equation}
Such a contribution like Eq.~\eqref{eq:sigma0-fermi-pinch-result}, where the fermion pinch singularity is incorporated, has been considered in  Section~\ref{sec:Resum-fermion-pinch}.

Next, when the product of the pair propagators in Eq.~\eqref{eq:sigma0-fp} leads to the pinch singularity \eqref{eq:pair-pinch}, it turns into
\begin{equation}
\begin{split}
&\sigma_{\fp}\vert _{\textrm{pair-pinch}}
= \frac{\beta\Omega_{d-1}}{m^{2}a^{4-d}}
 \int_{\bp,\bq}
 e^{-\beta\epsilon_{\bq}/2} \\
& \times
 \gamma^{F}(\epsilon_{\bq}/2-\epsilon_{\bq-\bp}-\epsilon_{B}-\mu+i0^{+},\epsilon_{\bq}/2-\epsilon_{\bq-\bp}-\epsilon_{B}-\mu-i0^{+};\bp) \\
&\times
 \calG^{R}(\epsilon_{\bq}/2-\epsilon_{\bq-\bp}-\epsilon_{B}-\mu,\bp)
 \calG^{A}(\epsilon_{\bq}/2-\epsilon_{\bq-\bp}-\epsilon_{B}-\mu,\bp)
 \frac{z^{2}e^{\beta\epsilon_{B}}\gamma^{B}_{RA\textrm{-pair}}(\bq)}
  {-\Im[\Delta^{R\textrm{-pair}}(\bq)]}
 +O(z^{2}).
\end{split}\xlabel{eq:sigma0-pair-pinch}
\end{equation}
In contrast to the previous case, the product of the fermion propagators cannot have the pinch singularity in this expression because $\calG^{R}(\epsilon_{\bp}/2+\epsilon_{\bq-\bp}-\epsilon_{B}-\mu,\bq)$ does not have the on-shell peak at $O(z^{0})$:
\begin{equation}
\Im[\calG^{R}(\epsilon_{\bq}/2-\epsilon_{\bq-\bp}-\epsilon_{B}-\mu,\bp)]
=\Im\left[\frac{1}{\epsilon_{\bq}/2-\epsilon_{\bq-\bp}-\epsilon_{\bp}-\epsilon_{B}+i0^{+}}\right]+O(z)=O(z)
\end{equation}
due to $\epsilon_{\bq}/2-\epsilon_{\bq-\bp}-\epsilon_{\bp}-\epsilon_{B}=-[2\epsilon_{\bq/2-\bp}+\epsilon_{B}]<0$.
The fermion propagators in Eq.~\eqref{eq:sigma0-pair-pinch} are simply replaced with the bare ones as
\begin{equation}
\calG^{R}(\epsilon_{\bq}/2-\epsilon_{\bq-\bp}-\epsilon_{B}-\mu,\bp)
\calG^{A}(\epsilon_{\bq}/2-\epsilon_{\bq-\bp}-\epsilon_{B}-\mu,\bp)
=\frac{1}{[2\epsilon_{\bp-\bq/2}+\epsilon_{B}]^{2}}+O(z),
\xlabel{eq:fermion-RA-on-binding-shell}
\end{equation}
and thus we obtain
\begin{equation}
\begin{split}
&\sigma_{\fp}\vert _{\textrm{pair-pinch}}
= \frac{\beta\Omega_{d-1}}{m^{2}a^{4-d}}
 \int_{\bp,\bq}
 e^{-\beta\epsilon_{\bp}/2}
 \frac{z^{2}e^{\beta\epsilon_{B}}\gamma^{B}_{RA\textrm{-pair}}(\bp)}
  {-\Im[\Delta^{R\textrm{-pair}}(\bp)]} \\
&\times
 \frac{\gamma^{F}(\epsilon_{\bp}/2-\epsilon_{\bp-\bq}-\epsilon_{B}-\mu+i0^{+},\epsilon_{\bp}/2-\epsilon_{\bp-\bq}-\epsilon_{B}-\mu-i0^{+};\bq)}
 {[2\epsilon_{\bq-\bp/2}+\epsilon_{B}]^{2}}
 +O(z^{2}),
\end{split}\xlabel{eq:sigma0-pair-pinch-result}
\end{equation}
with exchange of the integration variables.
In $\sigma_{\fp}$ with $\gamma^{B}(\ast)=\gamma^{F}(\ast)=O(z^{0})$, only Eqs.~\eqref{eq:sigma0-fermi-pinch-result} and \eqref{eq:sigma0-pair-pinch-result} give contributions at $O(z)$.
Therefore, we have shown that $\sigma_{\fp}$ has no leading contribution with the fermion and pair pinch singularities occurring simultaneously.

\section{Box diagram contribution for the bosonic vertex function}

\xlabel{app-sec:calculation-of-Box-vertex}
In this appendix, we derive Eq.~\eqref{eq:self-consistent-pair-vertex} by applying the analytic continuation of $i\omega^{F}_{m}\to \epsilon_{\bp}/2-\epsilon_{\bp-\bq}-\epsilon_{B}-\mu-i0^{+}$ followed by $i\omega^{B}\to i0^{+}$ to $\Gamma^{B}(i\omega^{B}_{m}+i\omega^{B},i\omega^{B}_m;\bp)$.
We take advantage of the fact that the box diagram with $\Gamma^{B}(\ast)$ can be regarded as a diagram in the right panel of  Fig.~\ref{fig:fermion-pair-four-pt-func}.
That is, when we take $\gamma^{B}(\ast)=m^{2}$ and
\begin{equation}
\begin{split}
&\gamma^{F}(i\omega^{F}_{m}+i\omega^{B},i\omega^{F}_{m};\bp) \\
&= \frac{2}{\beta}\sum_{l}\int_{\bk}
 G(i\omega^{B}_{l}-i\omega^{F}_{m},\bk-\bp)
 \calD(i\omega^{B}_{l}+i\omega^{B},\bk)\calD(i\omega^{B}_{l},\bk)
 \Gamma^{B}(i\omega^{B}_{l}+i\omega^{B},i\omega^{B}_{l};\bk)
\end{split}\xlabel{eq:rel-gammaF-GammaB}
\end{equation}
in Eq.~\eqref{eq:def-chi-fp-0}, $\chi_{\fp}(i\omega^{B})$ becomes the correlation function of the box diagram with $\Gamma^{B}(\ast)$.
Since $\chi_{\fp}(i\omega^{B})$ with the  pair pinch singularity leads to Eq.~\eqref{eq:sigma0-pair-pinch-result}, we can read off
\begin{equation}
\Gamma^{B}_{RA\textrm{-pair}}(\bp)
= m^2+\int_{\bq}
 \frac{\gamma^{F}(\epsilon_{\bp}/2-\epsilon_{\bp-\bq}-\epsilon_{B}-\mu+i0^{+},\epsilon_{\bp}/2-\epsilon_{\bp-\bq}-\epsilon_{B}-\mu-i0^{+};\bq)}
  {[2\epsilon_{\bq-\bp/2}+\epsilon_{B}]^{2}},
\xlabel{eq:GammaB-paring-shell}
\end{equation}
by comparing Eqs.~\eqref{eq:sigma0-pair-pinch-result} and \eqref{eq:pair-pinch-bulk}.
The vertex function~\eqref{eq:rel-gammaF-GammaB} is evaluated to the lowest order in fugacity as
\begin{equation}
\begin{split}
&\gamma^{F}(i\omega^{F}_{m}+i\omega^{B},i\omega^{F}_{m};\bp)
=2z\int_{\bk}e^{-\beta\epsilon_{\bk-\bp}}
 \calD(i\omega^{F}_{m}+i\omega^{B}+\epsilon_{\bk-\bp}-\mu,\bk) \\
&\quad\times
 \calD(i\omega^{F}_{m}+\epsilon_{\bk-\bp}-\mu,\bk)
 \Gamma^{B}(i\omega^{F}_{m}+i\omega^{B}+\epsilon_{\bk-\bp}-\mu,i\omega^{F}_{m}+\epsilon_{\bk-\bp}-\mu;\bk),
\end{split}
\end{equation}
which is dominated by the contribution from the pole of the fermion propagator because the branch cuts of the pair propagators contribute to $O(z^{2})$.
The analytic continuation of $i\omega^{F}_{m}\to \epsilon_{\bp}/2-\epsilon_{\bp-\bq}-\epsilon_{B}-\mu-i0^{+}$ followed by $i\omega^{B}\to i0^{+}$ leads to
\begin{equation}
\begin{split}
&\gamma^{F}(\epsilon_{\bp}/2-\epsilon_{\bp-\bq}-\epsilon_{B}-\mu+i0^{+},\epsilon_{\bp}/2-\epsilon_{\bp-\bq}-\epsilon_{B}-\mu-i0^{+};\bq) \\
& = 2z\frac{2\pi\Omega_{d-1}}{m^{2}a^{4-d}}
 \int_{\bk}e^{-\beta\epsilon_{\bk-\bq}}
 \frac{\delta(\epsilon_{\bp}/2-\epsilon_{\bp-\bq}+\epsilon_{\bk-\bq}-\epsilon_{\bk}/2)}{-\Im[\Delta^{R\textrm{-pair}}(\bk)]}
 \Gamma^{B}_{RA\textrm{-pair}}(\bk),
\end{split}\xlabel{eq:gammaF-pairing-shell}
\end{equation}
where the pair pinch singularity~\eqref{eq:pair-pinch} is applied.
Combining Eqs.~\eqref{eq:GammaB-paring-shell} and \eqref{eq:gammaF-pairing-shell} leads to Eq.~\eqref{eq:self-consistent-pair-vertex}.




\biboptions{sort&compress}
\bibliographystyle{elsarticle-num}

\bibliography{Bulk-viscosity-in-QVE}

\end{document}